\documentclass[structabstract]{aa}
\usepackage{graphicx}
\usepackage{txfonts}
\usepackage{natbib}
\bibpunct{(}{)}{;}{a}{}{,}

\begin{document}

\title{Evidence for a fast evolution of the UV luminosity function beyond redshift $6$ from a deep HAWK-I survey of the GOODS-S field.}
   \author{   M. Castellano \inst{1}
   \and
   A. Fontana \inst{1}
   \and
   K. Boutsia \inst{1}
   \and
   A. Grazian \inst{1}
   \and
   L. Pentericci \inst{1}
   \and
   R. Bouwens  \inst{2}
   \and
   M. Dickinson  \inst{3}
   \and
   M. Giavalisco  \inst{4}
   \and
   P. Santini \inst{1}
   \and
   S. Cristiani  \inst{5}
   \and
   F. Fiore \inst{1}
   \and
   S. Gallozzi \inst{1}
    \and
   E. Giallongo \inst{1}
   \and
   R. Maiolino  \inst{1}
   \and
   F. Mannucci  \inst{6}
   \and
   N. Menci \inst{1}
   \and
   A. Moorwood \inst{7}
   \and
   M. Nonino \inst{3}
   \and
   D. Paris \inst{1}
   \and
   A. Renzini \inst{8}
   \and
   P. Rosati \inst{7}
   \and
   S. Salimbeni \inst{5}
   \and
   V. Testa \inst{1}
   \and
   E. Vanzella  \inst{3}
   }

   \offprints{M. Castellano, \email{castellano@oa-roma.inaf.it}}

\institute{INAF - Osservatorio Astronomico di Roma, Via Frascati 33,
00040 Monteporzio (RM), Italy \and 
Lick Observatory, University of California, Santa Cruz, CA 95064, USA,\and NOAO, 950 N. Cherry Avenue, Tucson, AZ 85719, USA \and Department of Astronomy, University of Massachusetts, 710 North Pleasant Street, Amherst, MA 01003 \and INAF - Osservatorio Astronomico di Trieste, Via G.B.
Tiepolo 11, 34131 Trieste, Italy  
 \and INAF - Osservatorio Astrofisico di Arcetri, Largo E. Fermi 5, I-50125 Firenze, Italy   \and European Southern Observatory, Karl-Schwarzschild-Str. 2, D-85748 Garching, Germany \and INAF - Osservatorio Astronomico di Padova, Vicolo dell'Osservatorio 5, I-35122 Padova, Italy  }
   \date{Received .... ; accepted ....}

\titlerunning{Evidence for a fast evolution of the UV LF at $z>6$}
\authorrunning{M. Castellano et al.}

\abstract
{} { We perform a deep search for galaxies in the redshift range
  $6.5\le z\le 7.5$, to measure the evolution of the number density of
  luminous galaxies in this redshift range and derive useful
  constraints on the evolution of their Luminosity Function.  }
{ We present here the first results of an ESO Large Program, that
  exploits the unique combination of area and sensitivity provided in
  the near--IR by the camera Hawk-I at the VLT.  We have obtained two
  Hawk-I pointings on the GOODS South field for a total of $\sim32$
  observing hours, covering $\sim 90 ~ arcmin^2$. The images reach
  $Y=26.7$ mags for the two fields. We have used public ACS images in
  the $z$ band to select z-dropout galaxies with the colour criteria
  $Z-Y\ge 1$, $Y-J<1.5$ and $Y-K<2$. The other public data in the
  UBVRIJK bands are used to reject possible low redshift interlopers.
  The output has been compared with extensive Monte Carlo simulations
  to quantify the observational effects of our selection criteria as
  well as the effects of photometric errors.}
{We detect 7 high quality candidates in the magnitude range
  $Y=25.5-26.7$. This interval samples the critical range for $M_*$ at $z>6$   ($M_{1500}\simeq -19.5$ to $-21.5$).  After accounting for the expected incompleteness, we
  rule out at a 99\% confidence level a luminosity function constant
  from z=6 to z=7, even including the effects of cosmic variance. For
  galaxies brighter than $M_{1500}=-19.0$ we derive a luminosity density
  $\rho_{UV}= 1.5^{+2.0}_{-0.9}\times 10^{25} erg ~ s^{-1} ~ Hz^{-1} ~
  Mpc^{-3} $, implying a decrease by a factor 3.5 from $z=6$ to
  $z\simeq 6.8$. 
  On the
  basis of our findings, we make predictions for the surface densities
  expected in future surveys, based on ULTRA-VISTA, HST-WFC3 or JWST-NIRCam, evaluating the best observational strategy to maximise their
  impact.}  {}


\maketitle
%

\section{Introduction}

The search for extremely high redshift galaxies is much more than an
exciting exploration of the furthest frontiers of the Universe,
although this aspect is certainly at the basis of its popularity. Its
actual astrophysical interest is tied to the constraints that can be
set on the physical mechanisms that drove the formation and evolution
of galaxies at the earliest epochs of the Universe.

One important area of interest in the study of galaxies at $z>6$ is
ascertaining their role in the reionization of the Universe.  To be
fully responsible for the reionization, the density of star--forming
galaxies at $z>7$ should have been similar to that at $z\simeq 4$,
unless there is a significant evolution in the IMF and/or in the
clumpiness of the IGM, in the escape fraction of ionising photons or
in their metallicity \cite[see e.g.][and references
therein]{Mannucci2007,Oesch2009}.  Quantifying their number density is
therefore critical in order to constrain the additional mechanisms
that may be responsible for the re-ionizations, like Pop III dominated
primordial galaxies, mini-black holes or others \cite[see
e.g.][]{Venkatesan2003,Madau2004}.

Understanding the evolution of galaxies at high redshift is also very
important in the broader context of galaxy evolution.  While modelling
the growth of structures of dark matter is relatively straightforward,
modelling from first principles the physical mechanisms of star
formation and feedback that shaped galaxies across the life of the
Universe is remarkably complex.

The fundamental quantity that is currently used to describe and
quantify the galaxy population at high redshift is the UV Luminosity
Function (LF hereafter), as derived from surveys of Lyman break galaxies (LBGs) at
various redshifts. The evolution of its shape and normalisation along
the cosmic time provides a clear picture of the evolution of
star--forming galaxies in the early Universe, and an important
constraint on the related theoretical predictions.

Searches for LBGs have been extremely successful out to redshift 6 \cite[e.g.][]{Steidel1995,Steidel1999,Bunker2004,Dickinson2004,Giavalisco2004,Ouchi2004,Yoshida2006,Bouwens2003,Bouwens2006a,Bouwens2007,Mclure2009}
 i.e. when the Universe was only less than 1 Gyr old.  Current
samples of high-redshift galaxies ($z\sim3-6$) now contain tens of
thousands of galaxies and extend to luminosities as faint as $-16$ AB
mag (0.01 $L^*$).  

Despite these large samples, there is still controversy on how the UV
LF evolves at high-redshift.  Some authors
\citep{Sawicki2006,Iwata2007} have argued that the most significant
evolution in the UV LF happens at the faint end, others
\citep{Yoshida2006,Bouwens2006a,Bouwens2007} have found that the most
significant evolution is at the bright end, while \citet{Beckwith2006}
have claimed that the evolution is similar at the bright and faint
ends. Other authors have suggested compensating evolutions in LBG
number density and characteristic luminosity, resulting in a nearly
constant UV luminosity density \citep{Dickinson2004,Giavalisco2004}.
The likely origins of these discrepancies are both the large effect
of cosmic variance \cite[see e.g.][]{Trenti2008} as well as systematic effects due to the different
estimates of completeness level, contamination from lower redshift interlopers,  volume elements and redshift
distributions in the various samples \citep{Stanway2008b}, all worsened by the known
degeneracy among the parameters adopted to fit the LF. A clear example
of the effect of these uncertainties is shown by the recent revision
of the estimated slope of the UV LF at $z=2-3$, where LBG samples are
bigger and carefully characterised \citep{Reddy2009}.

Even within these uncertainties, it is becoming clear that the overall
evolution of the UV LF in the redshift range $z=2-6$ implies a
decrease in the number density of UV bright galaxies ($M <-20.5$) of a
factor 6-11 from $z\sim 3$ to $z\sim 6$
\cite[e.g.][]{Stanway2003,Shimasaku2005,Bouwens2006a}.

On the other hand, our knowledge of the evolution beyond $z\sim 6$ is
much more scanty.  Finding and studying galaxies at $z=7$ to constrain
the UV LF is definitely challenging, requiring deep and wide surveys
in the near--IR part of the spectrum.  Currently, the only constraints
come from a small sample of faint $z\sim7-9$ candidates found in small
areas within GOODS with deep near-IR $J+H$ NICMOS and WFC3 data
\citep{Bouwens2006b,Bouwens2008,Oesch2009,Bouwens2009b,Oesch2009b,Mclure2009b,Bunker2009}.
Dropout searches around lensing clusters have also been performed,
\citep{Richard2006,Richard2008,Bradley2008,Bouwens2009,Zheng2009}.
The discrepant results of these lensing studies clearly highlight the
difficulties in detecting $z>6$ candidates and in removing interlopers
from such samples \cite[see][for a discussion]{Bouwens2009}.

Further constraints come from the non-detection of bright candidates
of $z>6.5$ galaxies in relatively wide and shallower observation
\citep{Mannucci2007,Henry2009}.

Spectroscopic identifications of $z>6.5$ LBGs are lacking until now,
with the exception of the narrow band selected Ly$\alpha$ emitter at
z=6.96 by \citet{Iye2006}.

Apart from some contradictory results around lensing clusters, all the
evidences suggest that the number density of UV-bright galaxies fades
significantly at $z>6.5$, amounting to a decrease of the volume
density at the bright end of the UV LF of a factor 10-30 from $z\sim4$
to $z\sim7$ \citep{Mannucci2007,Bouwens2007,Stanway2008,Oesch2009}.

It is indeed possible to use these observations to
constrain the standard Schechter parameters of the LF, which appear
to deviate significantly from the $z=6$ ones, although the statistical
uncertainties are embarrassingly large. If one takes also into account
the many uncertainties due to systematic error in the candidate
selection and cosmic variance, it becomes clear that the evolution of
the LF at $z>6$ is still largely unexplored.

To progress in this field, larger and deeper IR--based surveys are
definitely needed. The very recent WFC3 data are providing a dramatic
advance, accessing the faint side of the LF at $z\simeq 7$ and
extending the searches to $z\simeq 10$
\citep{Bouwens2009b,Oesch2009b,Mclure2009b,Bunker2009,Yan2009}.  In parallel,
we are conducting a search of $z\simeq 7$ bright galaxies on wider
areas using the new VLT IR imager Hawk-I
\citep{Pirard2004,Casali2006,Kissler2008}, which is complementary to
it in many respects. First, we can cover areas significantly larger,
albeit shallower than WFC3, thanks to the wide field-of view of
Hawk-I. This will yield a statistically adequate sampling at the
brightest magnitudes, that is necessary to obtain an accurate estimate
of the LF parameters.

In addition, we will use the $Y$ band to detect galaxies at $z>6.5$, instead
of the $J$ band used so far. Thanks to the lower sky background and to the
extreme efficiency of Hawk-I, it is possible to reach the required
faint limits (close to $\simeq 27$ mags. AB). Such faint magnitudes can be
easily reached in the J band only from space.  More
important, the shorter central wavelength (around 1$\mu$m) of the $Y$ filter
corresponds to a lower and narrower redshift range, roughly $6.4<z<7.2$ (see
Section 4 and Figure\ref{simulnew}). This fair sampling is very important to
verify whether the UV LF evolves fastly at $z>6$.

In this paper we discuss the results of the first half of our survey,
covering a large fraction of the GOODS-S field. The paper is organised
as follows: in Section 2 we present our data set and the
multi-wavelength catalogue; in Section 3 we discuss our colour
selection criteria and the potential interlopers affecting the LBG
selection; in Section 4 we evaluate systematic effects in the light of
extensive Monte Carlo simulations; in Section 5 we present our final
sample of candidate z-drop Lyman Break Galaxies, which is used to
constrain the evolution of the $z>6$ UV luminosity function in Section
6. In Section 7 we discuss the implications of our findings, and of
present uncertainties on the high-z LF, on the efficiency of future
dedicated surveys. A summary of our methods and results is provided in
Section 8.

Throughout the whole paper, observed and rest--frame magnitudes are in
the AB system, and we adopt the $\Lambda$-CDM concordance model
($H_0=70km/s/Mpc$, $\Omega_M=0.3$ and $\Omega_{\Lambda}=0.7$).


\section{Data}

\subsection{The Data Set}
\label{dataset}

\begin{figure}
   \centering
   \includegraphics[width=9cm]{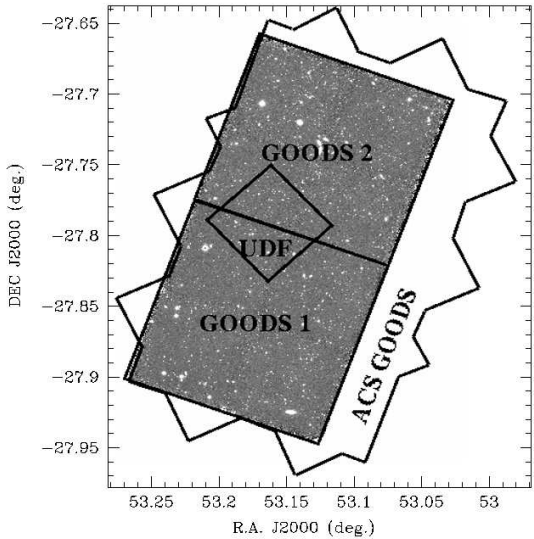}
   \caption{Full mosaic of the two Hawk-I images of the GOODS-S
     field. The jigsaw region shows the original GOODS ACS z-band
     image. The position of the UDF is also shown.}
         \label{image}
\end{figure}

This work is based on deep $Y$--band images obtained with Hawk-I, the
new near-IR camera installed at the VLT. With a field of view of about
7.5'$\times$7.5', pixel scale of $0.1"$, excellent sensitivity and image
quality, it is currently one of the best ground-based instrument to
search for faint, rare objects as very high redshift galaxies.

We combine data collected in 2007 during the Hawk-I Science
Verification phase with further images obtained in 2008 through a
dedicated ESO Large Programme. The images cover two adjacent regions
of the GOODS-S field, corresponding to $\sim 85\%$ of the deep ACS
area.  We name these two pointing GOODS1 (southern) and GOODS2 (northern).
The data obtained with the ESO LP extend the coverage of the GOODS2
field (originally about 11hrs in the $Y$ band) to the depth of the
GOODS1 area. The total exposure time is 16h15m for GOODS1 and 16h56m
for GOODS2. The position of the two combined Hawk-I fields is shown in
Figure~\ref{image}.

The $Y$ band images were reduced  using standard techniques
for IR data - flat fielding, sky subtraction among consecutive frames,
and final coaddition. Particular care was taken in masking
even the faintest sources during the sky subtraction. This has been
accomplished with a ``double-pass'' procedure, where the faintest
sources have been detected in a first version of the finally coadded
images, and the whole data reduction has been repeated  masking all
objects during the sky estimate step.

Two classes of defects are present in the Hawk-I images. One class
consists of luminous ``ripples'' along the rows originating from bright
saturated objects. These defects are due to cross-talk effects and
have been removed by masking the whole range of affected rows in each
image. Further problems are caused by a ``persistence'' effect, i.e.
residuals left by sources which appear as faint objects on the next
acquisition at the previous position (in pixel) of the bright
object. These ``ghosts'' may appear as faint dropouts in the final
image, unless some masking is done. We have decided to remove these
objects from the beginning, by masking (in each image) the pixels
where objects were detected in the previous one ($\sim$0.3-1.0\% of the total
area in each science frame). This strategy turned
out to be more effective and safer than just making a sigma-clipping
during the final coaddition.

The final images have been registered to the ACS astrometric solution,
maintaining the original Hawk-I pixel size.

We find that the GOODS1 and GOODS2 fields are rather uniform in their
general properties.  Through the analysis of bright point-like sources
we determine a FWHM of 0.51 $\pm 0.01$ arcsec (= 4.8 pixels) in the GOODS1 image
and 0.49 $\pm 0.02$ arcsec (= 4.6 pixels) in the GOODS2 one.  We computed image
zeropoints using the standard stars observed during the same night of
the GOODS data and at similar airmasses. Since the standard stars are
calibrated on A0V stars in the MKO filter set, we have estimated the
conversion to the AB system for the Hawk-I specific $Y$ band filter
using templates, as $Y_{Hawk-I}=Y_{MKO} + (J_{MKO}-K_{MKO})$. The
resulting zeropoints are $Y=26.992$ and $Y=26.998$ (AB), for GOODS1
and GOODS2 respectively.

We have also carefully determined  the r.m.s. of the coadded images,
which is necessary to properly estimate the statistical meaning of
detections at the faint limit. We have obtained ``from first
principles'' an absolute r.m.s. which fully accounts for the
correlation in the pixel of the finally coadded image. This was
estimated by computing the r.m.s. in each individual image (using the
Poisson statistics and the instrumental gain) and propagating
self-consistently this r.m.s. over the whole data reduction process.
The typical $5\sigma$ magnitude in one arcsec$^2$ is in the range
26.7-26.8 over more than 60\% of the whole image, and $>26.2$ in 85\%
of the image - the rest of the images being shallower because of the
gaps between the four Hawk-I chips.

\subsection{The Photometric Catalog - Detection}\label{detection}
We have obtained the photometric catalog using the SExtractor code
\citep{Bertin1996} and the $Y$ band as detection image. The
r.m.s. derived ``from first principles'' as described above is used in
SExtractor as \verb|MAP_RMS| and overrides the r.m.s. obtained by
SExtractor from the background fluctuations.  To obtain total
magnitudes, we have computed both the SExtractor's \verb|MAG_BEST| as
well as aperture corrected magnitudes in 2 FWHM diameter (about 1''),
the same apertures used for the colour estimate discussed below.  We
compute the aperture corrections from bright non-saturated stars in
each field: we find corrections of 0.364 mags and 0.322 mags in GOODS1
and GOODS2, respectively. \verb|MAG_BEST| magnitudes are more accurate
for bright objects, but become fainter than aperture--corrected ones
at about $m_Y\simeq 24$.  As we shall show below, all $z>6.5$
candidates are so faint that their total magnitude is estimated with
aperture-corrected magnitudes. For resolved objects, however, aperture
corrections based on stellar profile may underestimate the actual
total flux.  We have estimated this effect by using Lyman Break
Galaxies with known spectroscopic redshifts $5.5<z<6.2$ in the GOODS-S
ACS images \citep{Vanzella2009}, smoothing them to the Hawk-I PSF. We
find that the aperture correction is slightly larger, i.e.  $\sim
0.5$.  For simplicity, and given the unknown physical extent of our
$z>6.5$ galaxies, we shall adopt the nominal aperture correction based
on known stars for all candidates. However, the effect of lost flux
due to finite size of our candidates is fully accounted for by the
simulations that we use to estimate the Luminosity Function, as we
shall discuss in Section \ref{simulations}.

The most critical issue is the very detection at faint limits. Given
the extreme faintness expected for the $z>6.5$ galaxies and their
rarity, a compromise must be found between two competing goals:
extending the detection at the faintest possible levels while
retaining a good accuracy. In addition, the systematics must be
understood and quantified for the final scientific analysis. We have
optimised the SExtractor parameters involved in the detection of faint
objects evaluating at the same time the possible contamination from
spurious objects in the Y-detected catalogue through the analysis of a
'negative' image. To do this, we vary the detection parameters used by
SExtractor (\verb|DETECT_MINAREA|, \verb|DETECT_THRESH|, filtering,
deblending and background subtraction parameters) in the analysis of
the 'positive' image. We then construct a ``background-subtracted
negative'' image and analyse it exactly as the positive one, with the
same SExtractor parameters. The RMS image used is the same employed in
the analysis of the 'positive' image. We finally adopt the set of
parameters that minimises the ratio between 'negative' and 'positive'
detections at the faint end of the number counts. The final detection
is obtained requiring 10 contiguous pixels each at $S/N>0.727$,
corresponding to a $2.3\sigma$ detection, and restricting the analysis
to the regions where the R.M.S is less than 1.5 times the lowest
value.  With this choice of parameters, detections on the negative
images are negligible at $Y<26.5$, and less than 10\% down to $Y=26.8$
in both fields. However, \textit{a posteriori}, the latter
  value overestimates the actual rate of spurious detections. Indeed,
  all spurious sources should appear as ``drop-out'' candidates with a
  single-band detection. On the contrary, as we will show in
  Sect~\ref{7galaxies}, our two faintest candidates 
  are both confirmed by detections in other IR
  bands. A visual inspection of the negative images shows that many
  faint 'objects' are found near bright sources: this is an indication
  that, at the faintest limits, non trivial issues concerning the
  subtraction of the background or a potential asymmetry in the noise
  distribution produce an overestimate of the rate of spurious
  detections.  

This procedure is also used to define the total area where a homogeneous
catalog can be extracted. The candidates found in this area will be
used for the evaluation of the LF. This area is $\sim 75$\% of the
whole image and is $89.7$arcmin$^2$. We remark that additional
candidates satisfying our colour and S/N selection criteria (see Sect. \ref{Selection}) are also found in the more noisy regions of the
images, the most notable being an object falling in the UDF, but they will
not be discussed in this paper. 

\subsection{The Multicolour Catalog}\label{catal}
We have obtained full multiwavelength photometry of all the objects
detected on the $Y$ band using the publicly available UBVRIZJHK
images. ACS BVIZ images are the latest V2.0 version released by the
STSci (M. Giavalisco and the GOODS Team, in preparation).  Each frame
has been smoothed to the Hawk-I PSF using an appropriate kernel
obtained with Fourier transform \citep{Grazian2006} and registered to
the Hawk-I images. Publicly available U and R images obtained with
VIMOS \citep{Nonino2009} and JHK mosaics (Retzlaff et al., in
preparation) have also been registered to the Hawk-I images. These
have not been filtered, since their PSF is always somewhat larger than
the $Y$ images. We then compute magnitudes in $U$, $B$, $V$, $R$, $I$,
$z$, $J$, $H$ and $Ks$ bands running SExtractor in dual mode using the
Y band HAWK-I image as detection image with the detection parameters
indicated above. Aperture fluxes are computed with the same aperture
as in the Y band, and appropriate aperture corrections have been
separately applied to each band.  A comparison between ACS and ISAAC
total magnitudes in our catalogue and in the GOODS--MUSIC catalogue
\citep{Grazian2006,Santini2009} shows a good agreement, apart from a
fraction of blended objects (10\% of the total) that are unresolved in
the Y band detection but have been deblended in the ACS $Z$-detected
sample of the GOODS-MUSIC catalogue.  The typical $1\sigma$ limiting
magnitudes in these images, scaled to the total $Y$ band flux of
detected objects (i.e. estimated in the 1.2'' aperture and corrected
to total) are 29.1, 29.0, 29.1, 29.5, 28.6, 28.2, 26.6, 26.2, 26.2 in
U,B,V,R,I,Z,J,H,K respectively.

This catalog contains self-consistent magnitudes in all bands.
To exploit the superior image quality of the ACS images, we
have also obtained  photometry of all objects on the BVIZ images
without smoothing them to the Hawk-I PSF, in a narrower aperture of
0.6'' (an even smaller aperture would be prone to errors on object
centering).

\section{The Selection of z$>6.5$ Galaxies}\label{Selection}
\subsection{The Colour Selection Criterion}

\begin{figure}
   \centering
   \includegraphics[width=8cm]{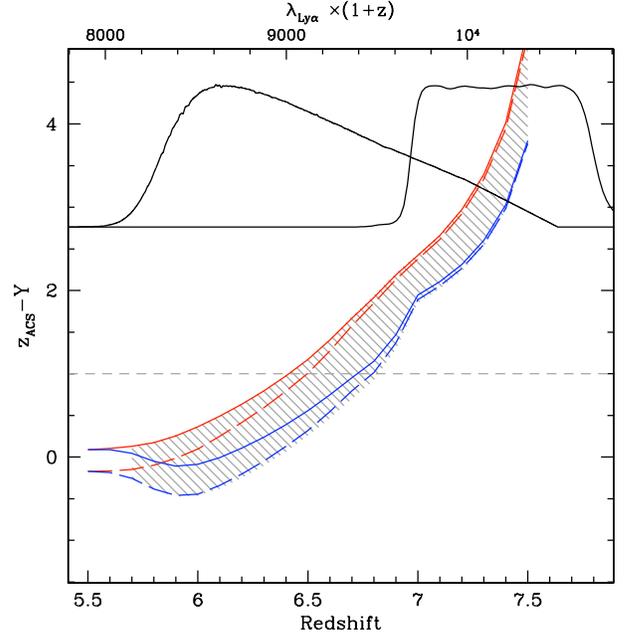}
   \caption{$Z-Y$ colour as a function of redshift in the Hawk-I filter
     set. Shaded area shows the locus predicted by CB07 models with a
     range of metallicities, ages, dust extinction and Lyman-$\alpha$
     emission (see text for details). Red lines corresponds to extreme
     models with no Lyman-$\alpha$, blue lines with a rest frame
     Lyman-$\alpha$ equivalent width of $300$\AA. In the upper part,
     the efficiency curve of the two filters is shown, computed at
     observed wavelength of a Lyman-$\alpha$ emission at the
     corresponding redshift.}

         \label{Z_Y}
\end{figure}

The selection of galaxies at $z>6.5$ uses the well known ``drop-out''
or ``Lyman-break'' technique, with minor modifications due to our
filter set and imaging depth. The main spectral feature that enables
the identification of galaxies at extreme redshifts is the sharp drop
shortward of the Lyman-$\alpha$, where most of the photons are absorbed
by the intervening HI in the intergalactic medium. At $6.5<z<7.5$,
this break is sampled by the large $Z-Y$ colour, as shown in
Figure~\ref{Z_Y}, where we plot the $Z-Y$ colour of galaxies at $z>6$
in our filter set. At this purpose we have used the models of Charlot
and Bruzual 2007 \cite[in preparation, see][hereafter
CB07]{Bruzual2007a,Bruzual2007b} with the following range of free
parameters: Metallicity: 0.02, 0.2 and 1 $Z_\odot$; age from 0.01 Gyr
to the maximal age of the Universe at a given $z$; E(B-V)=0...0.2
\citep{Calzetti2000}. Since Lyman-$\alpha$ emission has an
  important influence on the selection function of Lyman break galaxies, as it has been proved in the case of lower redshift samples
  \cite[e.g.][]{Stanway2008b,Dow2007}, we explicitely take into account this
  effect by considering a distribution of Ly-$\alpha$ rest-frame
equivalent width in the range 0-300 \AA. 
We have also added the intergalactic absorption
using the average evolution as in \citet{Madau1995}. The shaded area
shows the region covered by the computed models.  It is evident that
the colour evolution is relatively smooth, mainly because of the
extended red tail of the $Z$ filter which gathers a fraction of
flux from galaxies up to $z\simeq 7.5$.

It is interesting to note that most of the broadening in the $Z-Y$
colour distribution of the model galaxies in Figure~\ref{Z_Y} is due to
the effect of the Lyman-$\alpha$ emission line, which may change the
observed colour (at a given redshift) by nearly $Z-Y\simeq 1$
mags. 
The uncertainty in the intensity of this line can be translated into
an uncertainty in the estimated redshift, which is (even ignoring
noise effects) about $\Delta z \simeq 0.2 - 0.4$ in the redshift range
$6.5<z<7.5$. Despite this, we shall adopt a single
threshold $Z-Y\geq 1$ to select candidates. 

\begin{figure}
   \centering
   \includegraphics[width=8cm]{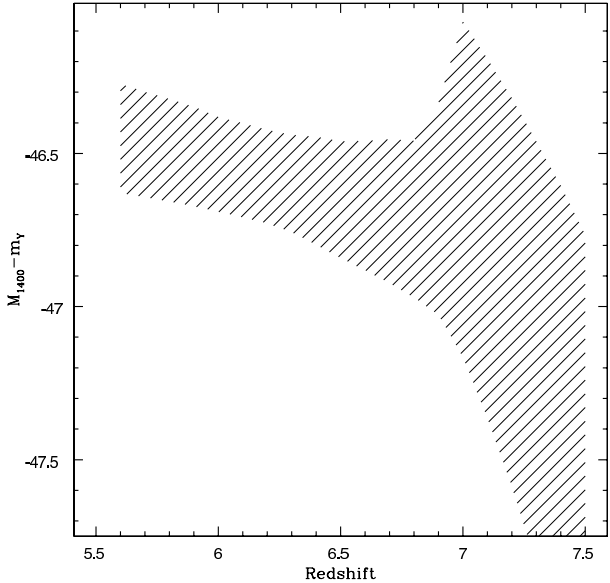}
   \caption{The  distance modulus $M_{1500} - m_Y$ as a function of
     redshift in the $Y$--band Hawk-I filter. Shaded area shows the
     locus predicted by CB07 models with a range of metallicities,
     ages, dust extinction and Lyman-$\alpha$ emission (see text for
     details).  }
         \label{DM}
\end{figure}

The major difference with the standard Lyman break technique is that
the $Y$ band that we use to detect galaxies does not sample the
continuum around $1500$ \AA ~ but a region shortward of it,
contaminated by both the IGM absorption as well as by the Lyman-$\alpha$ emission line.  Since we aim at comparing our
candidates with those expected from the LF at 1500
\AA, this effect can be taken into account by computing the expected
distance modulus $DM(z) =  M_{1500} - m_Y$ with the same template set
described above.  This is shown in Figure~\ref{DM}. In addition to the
cosmological dimming with redshift (for galaxies of given $M_{1500}$),
the Lyman-$\alpha$ emission can produce a brightening at $z\geq 7$,
rapidly counterbalanced at $z\geq 7.3$ by a loss of flux due to the
intervening IGM absorption.

All these effects will be accounted for by the Monte Carlo treatment
that we shall discuss in the next Section.

\subsection{Interlopers}

\begin{figure}
   \centering
   \includegraphics[width=8cm]{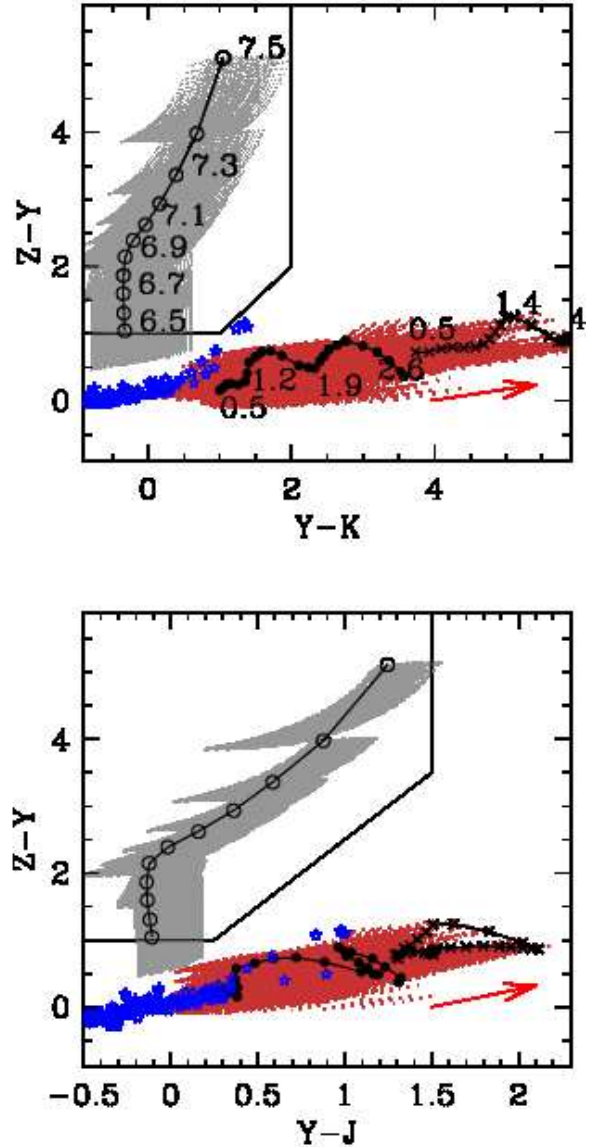}
   \caption{{\it Upper panel}: $Z-Y$ - $Y-K$ diagram showing the expected position of
     $z>6.5$ galaxies (grey dots), passively evolving galaxies and
     reddened starbursts (red dots) at different redshifts. See text for the
     details of the adopted library. The redshift evolution of a single
     representative model is also shown: open circles for LBGs, large filled
     circles for passively evolving, crosses for dusty starbursts. The red
     arrow  represents the reddening vector at $z=1.5$. Blue
     symbols mark the position of normal galactic stars. 
  {\it Lower panel}: as above, for the $Z-Y$ - $Y-J$ colour plane. 
   }
         \label{interlopers}
\end{figure}

Given the rarity of high redshift galaxies, it is mandatory to discuss
the possible contaminants in the selected colour criterion.

Apart from $z>6$ galaxies, other well known classes of objects can
display a red $Z-Y$ colour: i) Variable objects, mostly due to
low-intermediate redshift SN events; ii) Passively evolving galaxies
or dusty starburst galaxies at $z>1.5$; iii) Galactic cool stars.

We have first cross-checked each object with $Z-Y>1$ against
variability, by looking at images acquired at different epochs (1
month of delay in GOODS1, 1 year in GOODS2). We identified three
obvious transients in the GOODS1 images, that have been removed from
the following analysis. All the other objects in our sample have a
consistent photometry in the two epochs, at the $3\sigma$ level.

Passively evolving galaxies or dusty starburst galaxies at $z>1.5$ can be
easily modelled with a suitable set of spectral synthesis models. We
use the same CB07 library used for $z>5.5$ galaxies to predict the
colours of such objects at $1.5<z<4$, using a combination of short
star-formation exponential timescales ($0.1-1$ Gyrs) and ages $>1$ Gyr
to reproduce passively evolving galaxies, and constant star-forming
models with $0.5<E(B-V)<1.5$ \cite[adopting a ][extinction law]{Calzetti2000} 
for the dusty starbursts.  As shown in Fig.\ref{interlopers}, these
galaxies may have large $Z-Y$ colours only when they also show large IR
colour terms in the J and K bands. We note that the effect of even larger
amount of dust extinction would be to shift the objects at even redder $Y-J$
or $Y-K$ colors. This is shown by the red arrow in Fig.~\ref{interlopers}, which
represents the reddening vector at $z=1.5$. The slope of the reddening vector
at other redshifts is very similar.
To exclude these objects, we shall
adopt in the following the additional criteria (see Fig.~\ref{interlopers}):

\begin{eqnarray*}
(Z-Y) &>& 1.0+(Y-K)\\
(Z-Y) &>& 0.5+2.0(Y-J)\\ (Y-J)&<& 1.5 \\(Y-K)&<& 2.0
\end{eqnarray*}
It is less straightforward to exclude T-dwarfs. They are cool
($T_{eff}<1500$ K), low-mass stars and substellar objects that have
been discovered and studied in these last ten years mainly thanks to
the 2MASS \cite[e.g.][]{Burgasser1999}, SDSS
\cite[e.g.][]{Leggett2000}, UKIDSS \cite[e.g.][]{Pinfield2008} and
CFBDS surveys \citep{Delorme2008}. Their infrared spectra are
dominated by the presence of $CH_{4}$ and $H_2O$ absorption bands and
by $H_{2}$ resonant absorption
\cite[e.g.][]{Chabrier2005,Burgasser2006} that produce a sharp break in
their IR colours which resembles the $Z-Y$ and $Z-J$ colours of high
redshift LBGs. Moreover faint dwarfs can easily be beyond the
detection limit even in the I band because of their cold atmospheric
temperatures.

Unfortunately, the current uncertainties in cool dwarfs atmospheric
models \citep{Helling2008}, do not allow us to compute the expected
$Y-J$, $Y-H$ or $Y-K$ colours with the reliability needed to
distinguish T-dwarfs from LBGs. In addition, such distinction is
unfeasible on the basis of their morphology/size in our ground-based Y
band images, since morphological classification is not reliable at
very low S/N ratios.

It is however possible to estimate their expected number.  While cool
dwarfs are found also at magnitudes much brighter than those of $z>6$
galaxies, their number density is known to increase towards fainter
fluxes \citep{Dantona1999,Burgasser2004}, so it is not possible to
exclude that some of these objects contaminate our sample of high
redshift candidates.  The exact number of expected contaminants
depends on the still uncertain parameters constraining the IMF and the
spatial distribution of late type dwarfs inside the disk and the halo
of the Galaxy. Observations have excluded that these objects can be
described by a power-law IMF$~\Phi(M) \propto M^{-\alpha}$ as steep as
the Salpeter one ($\alpha=2.35$), with an upper limit at $\alpha=1.5$
\citep{Burgasser2004}.  For this reason we considered a worst-case
model with an IMF exponent $\alpha=1.5$, and a spatial distribution
with height of the Galactic disk $H_z=300 \ pc$, following the
T-dwarfs surface density predictions presented by
\citet{Burgasser2004} for a deep survey at high Galactic latitude, we
obtain an estimated contamination of $\sim 1.7$ late type dwarfs
($T_{eff}<1750 K$) for each Hawk-I pointing. Larger values for the
 T-dwarfs scale height in the disk \cite[as the 350 pc value proposed by][]{Ryan2005}, or an even
 steeper IMF, could slightly increase this already pessimistic
 estimate. However, we underline here that, with such a low number of expected
 cool dwarfs, the main uncertainties arise from poissonian fluctuations in their
 number counts.  We note that two cool
dwarfs with compatible colours were indeed found in the GOODS field by
\citet{Mannucci2007}. They both have $z-Y>1$ in our catalogue: one
(ID=6968 in \citet{Mannucci2007}, G1\_1713 in our catalog having
Y=24.45) is selected as candidate drop-out, while the other (ID=4419
in \citet{Mannucci2007}, G1\_5130 in our catalog, with Y=23.3) is
rejected because it has a significant detection also in the I band.

Apart from these well know classes of possible interlopers we found
that a sample of galaxies selected with the $Z-Y>1$ and $Y-K<1$
criterion is populated also by an unexpectedly large number of
faint contaminants showing significant detection in filters covering
wavelengths shorter than the redshifted Lyman limit at $z>6$ (U and R
VIMOS; B, V, I ACS).

\begin{figure}
   \centering
   \includegraphics[width=8cm]{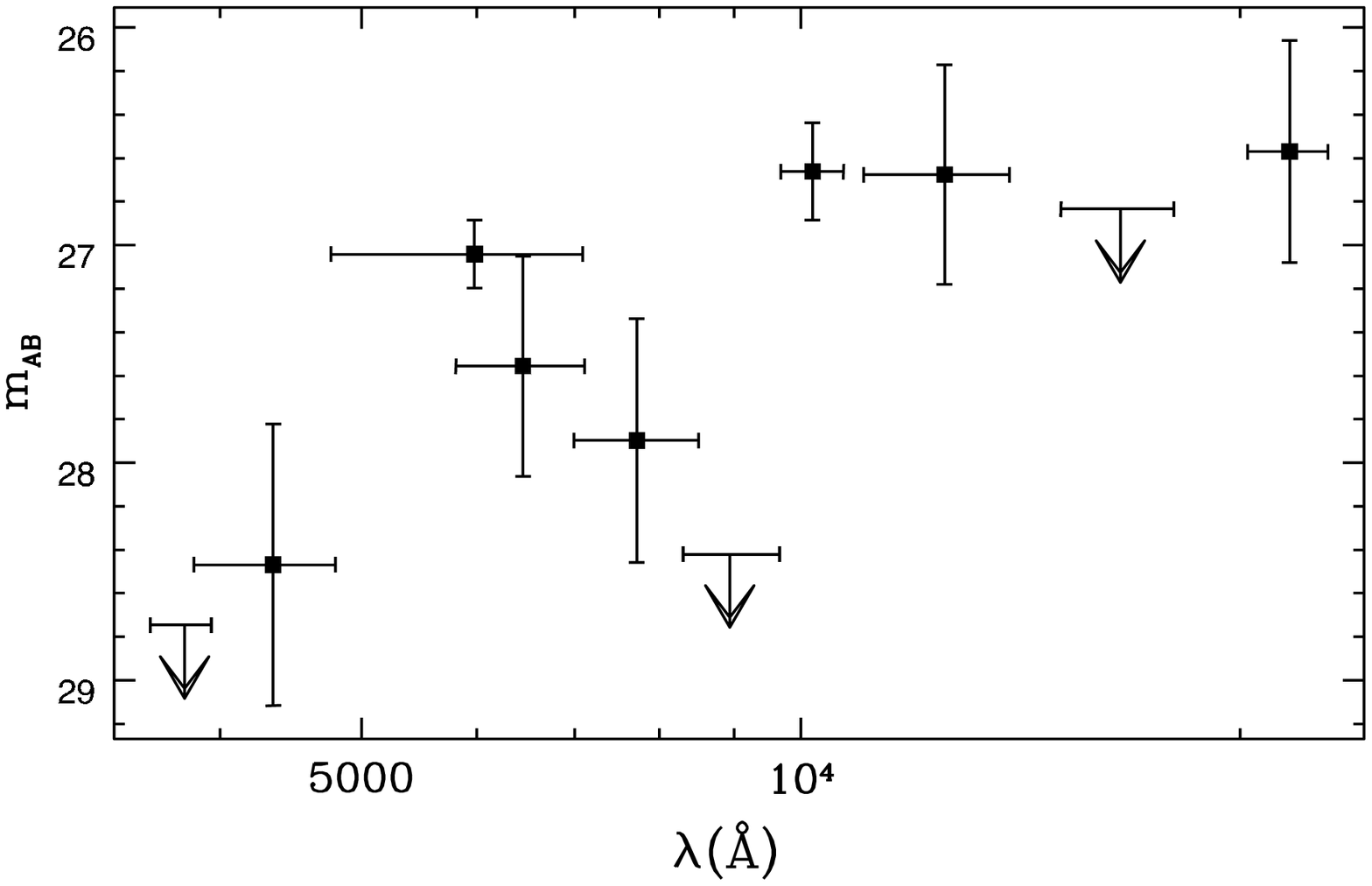}
   \includegraphics[width=8cm]{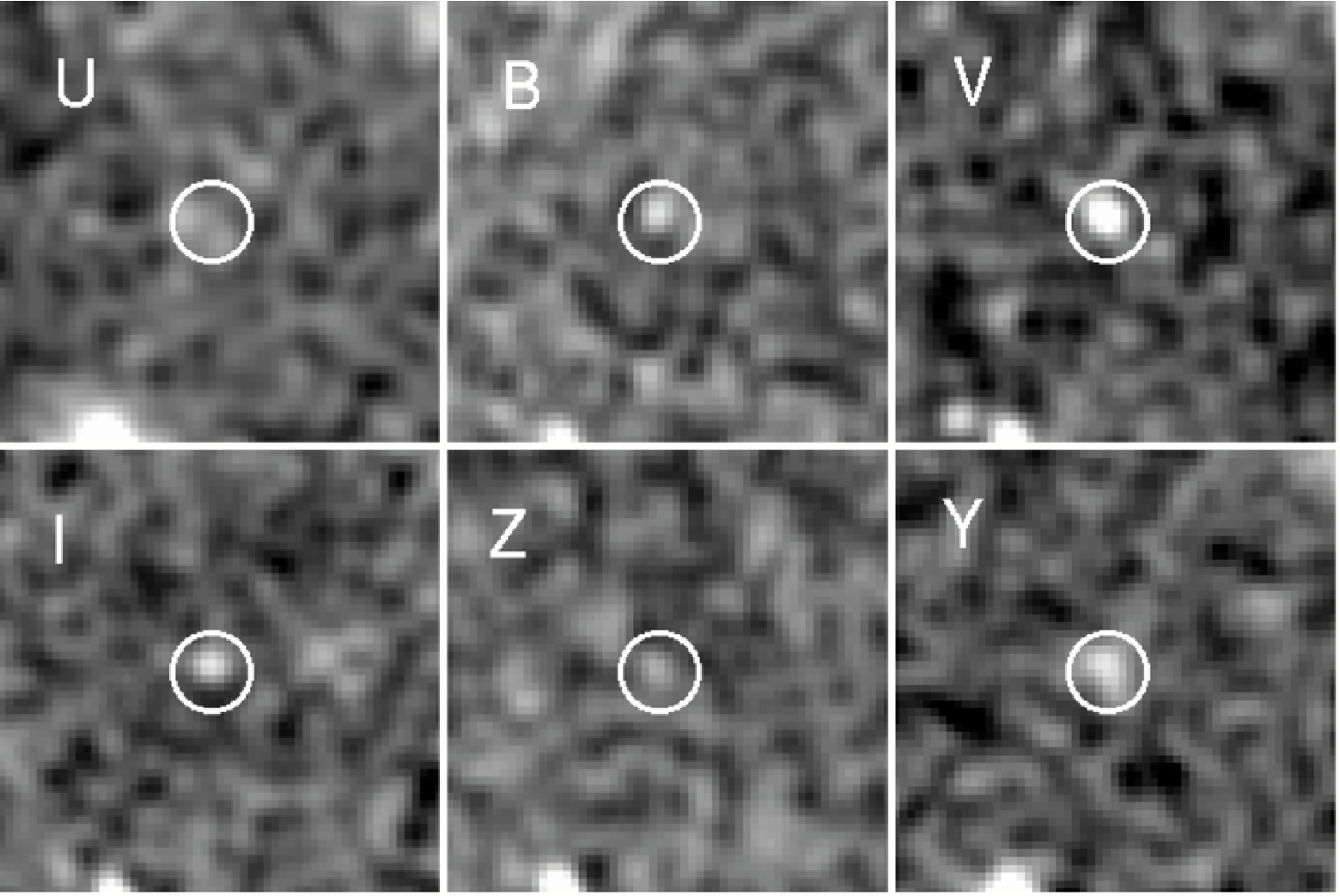}
   \caption{SED and thumbnail for one of the interlopers with $Z-Y>1$
     and detection at shorter wavelengths.}
         \label{contam}
\end{figure}

We show in Fig. \ref{contam} the SED and the image in different
filters of a typical member of this contaminating population. These
objects show a dip in the observed flux in the $Z$ and often in the
$I$ band, with a rising continuum in the bluer bands and a large $Z-Y$
colour. Such a spectral energy distribution cannot be reproduced by a
straightforward application of the CB07 models. The analysis of the
nature of these objects is beyond the scope of the present paper and
will be carried out separately. We only note that all of them are
relatively faint (mostly $m_Y>25$). A visual inspection of
  their ACS images revealed that these objects show a variety of morphologies, 
  at least those detected at good
  S/N. In many cases they are clearly extended, thus indicating their
  extragalactic nature, although other cases present a point-like
  morphology. One possibility, 
 especially for those
undetected in $JHK$, is that they are faint galaxies with a very blue
continuum whose SED is altered by the presence of strong emission
lines such as in un-obscured AGN, or in star forming galaxies like the
blue compact dwarf galaxies \citep{Izotov2004,Izotov2007} or the Ultra
Strong Emission Line Galaxies \cite[USELs,][]{Hu2009}.

Although standard prescriptions for the selection of high redshift
LBGs already exclude sources detected at short wavelengths, given the
rarity of $z>6$ candidates we took particular attention in tailoring
reliable criterion to separate our sample of high redshift candidates
from these unexpected lower redshift objects.  After looking at the
colour distributions of these contaminants and at the distribution of
the detected S/N in the UBVRI bands, we decided to adopt a selection
criterion even more conservative that the one typically adopted by
previous surveys. We require the S/N from high-z candidates to be
$<2\sigma_{S/N}$ in all UBVRI bands and $<1\sigma_{S/N}$ in at least
four of them. $\sigma_{S/N}$ is the estimated r.m.s. of the S/N
distribution in each band, as described in the next Section. In the
case of the ACS BVI images we used S/N ratios, and relevant
$\sigma_{S/N}$, measured in a smaller aperture (0.6'', see
Sect. \ref{catal}) to better exploit their higher resolution.
To check whether our selection criteria are effective in
  excluding this population of faint contaminants from the LBG sample,
  we have also analysed the deep ($\sim$ 30 mags AB at 1$\sigma$)
  public images of the UDF area \citep{Beckwith2006}, which is also
  covered by our shallower data-set (see Fig.~\ref{image}).  Those
  objects that, on the basis of the conservative criteria discussed
  above, are individuated as contaminants in our catalogue, are all
  confirmed, at an higher significancy, to have emission in one or
  more of the blue bands.  On the other hand, we effectively select
  as an high redshift candidate  object UDF-387-1125 already discussed in
  \citet{Bouwens2004} and recently confirmed in the analysis of the very
  deep WFC3 images by \citet{Bunker2009,Mclure2009b,Oesch2009b,Yan2009}. 
  Thus we can state that
  deep optical images as the ones analysed here ($\sim$ 29.0 mags
  AB at 1$\sigma$) are sufficient in separating this class of
  contaminants from more reliable high redshift candidates, once
  conservative selection criteria are adopted.  However, we caution
  that an accurate, dedicated, spectroscopic analysis of both LBGs and
  contaminants will be necessary to determine the real impact of these
  objects in LBG searches, since their physical nature is still
  undetermined.  

\section{Simulating the Systematic Effects}
\label{simulations}

\begin{figure}
   \centering
   \includegraphics[width=8cm]{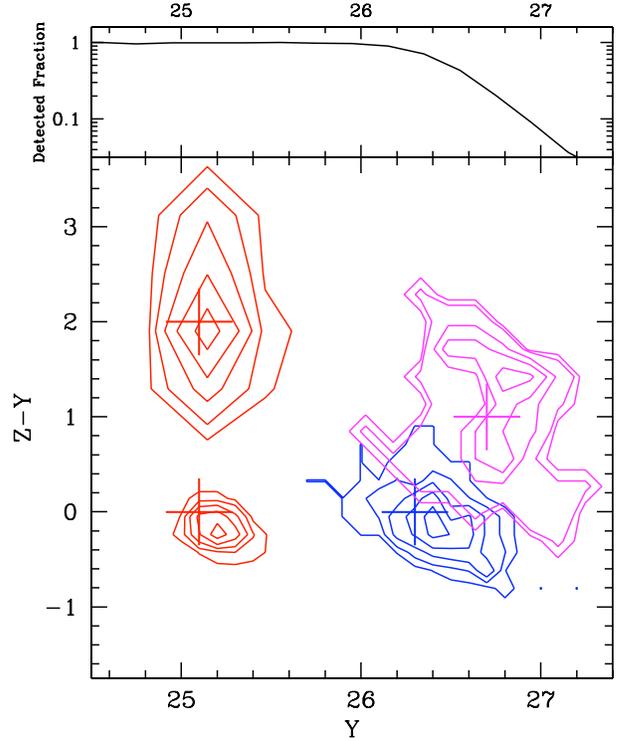}
   \caption{{\it Upper panel}:Fraction of detected objects as a
     function of their input magnitude, as estimated from the
     simulations described in the text. {\it Lower panel}: From the
     same simulations, contour levels of the conditional probability
     $P(Y_M,~(Z-Y)_M ~|~ Y,~(Z-Y)~)$ that a galaxy with given $Y$ band
     flux and $Z-Y$ colour is detected with a measured $Y_M$ flux and
     with a measured $(Z-Y)_M$ colour. Only four cases are shown, with
     input values marked by large crosses. Contours are drawn at 0.05, 0.15,
     0.3, 0.5 and 0.8  times the peak value of each distribution.}
         \label{simul_mat}
\end{figure}

Although the colour criteria are formally very clear-cut, they are in
practice applied on very faint objects, typically close to the
limiting depth of the images. At these limits, systematics may
significantly affect their detection and the accurate estimate of their
large colour terms.  To fully evaluate the involved uncertainties we
have performed extensive imaging simulations, by which we have
quantified the systematic effects in the object detection, in the
measure of their total magnitude as well as in the measure of large
colour terms in our images. The output of these simulations will be
used in the proper estimate of the LF.

At this purpose, we first use the synthetic libraries described above
to produce a large set of simulated galaxies with expected magnitudes
in the UBVRIZY filter set in the redshift range $5.5<z<8$. Objects are
normalised in the range $Y=24-27.5$, following the expected magnitude
distribution arising from a LF with index $\alpha=-1.71$, These
galaxies are placed at random positions of the GOODS1 and GOODS2
$Y$-band images, and catalogs are extracted exactly as in the original
frames. 

As expected, the output of this exercise depends critically on the
assumed morphology. In our case, the availability of deep $z$-band ACS
images of confirmed $z=5.5-6$ LBGs provides the most natural
templates, avoiding further assumptions.  Neglecting possible size
evolution from $z=6$ to $z=7$, we have used the four brightest
$z=5.5-6$ LBGs observed with ACS in GOODS, both convolved with the
GOODS1/2 PSFs.

To avoid an excessive and unphysical crowding in the simulated images,
we have included only 200 objects of the same flux and morphology each
time, after masking the regions of the images where real objects have
been detected (this corresponds to 6\% of the total area). We repeated
the simulation until a total of $10^5$ objects were tested for each
field and morphology, for a total of $8\times10^5$ artificial objects.

These simulations can be used to estimate the impact of different
systematics in the various steps of the analysis.  First, they can be
used to estimate the completeness in the detection procedure. This is
shown in the upper panel of Figure~\ref{simul_mat}, where we plot the
fraction of detected versus input objects as a function of the input
magnitude.  The detection is close to 100\% down to $Y\simeq 26.2$,
and fades to 30\% at $Y\simeq 26.7$ (where we find our fainter
candidates. We remind that an additional 6\% in the incompleteness is
due to the masking done to avoid bright objects.

In addition, they can be used to evaluate the uncertainties in the
estimate of the colour criteria that we shall exploit to detect $z>6.5$
candidates. In particular, the $Z-Y$ colour is critical feature to
identify $z>6.5$ objects. To visualise the effect of noise, we plot
in the lower panel of Figure~\ref{simul_mat} the conditional
probability $P(Y_M,~(Z-Y)_M ~|~ Y,~(Z-Y)~)$ that a galaxy with given
$Y$ band flux and $Z-Y$ colour is detected with a measured $Y_M$ flux
and with a measured $(Z-Y)_M$ colour. Obviously, a fraction of
galaxies will not be detected at all in the $Y$ band, as discussed
above.

An inspection of Figure~\ref{simul_mat} shows the basic feature of the
systematics acting on our images. At relatively high S/N ($Y\simeq
25$, $Z\simeq 25$), the recovered magnitudes and colours have a narrow
scatter, $\sigma_Y\simeq \sigma_z\simeq 0.07$ mags which increases at
lower fluxes ( $\sigma_Y\simeq \sigma_z\simeq 0.2$ mags at $Y\simeq
26$, $Z\simeq 26$). At larger $Z-Y$ colours asymmetries becomes evident
as the $Z$ magnitude approaches the very detection limit. At the
faintest limit, ($Y\simeq 26.8$ and $Z-Y\simeq 1$), about 30\% of the
objects detected in $Y$ become too faint to be detected in the $Z$
band image.  

The simulations in Figure~\ref{simul_mat} also show that a 0.15
mags offset exists on average between the input and the recovered
magnitude. As discussed above, this is due to the adopted aperture
correction, which is computed on unresolved stellar profiles, instead
of the LBG profile adopted in the simulations.

\begin{figure}
   \centering
   \includegraphics[width=8cm]{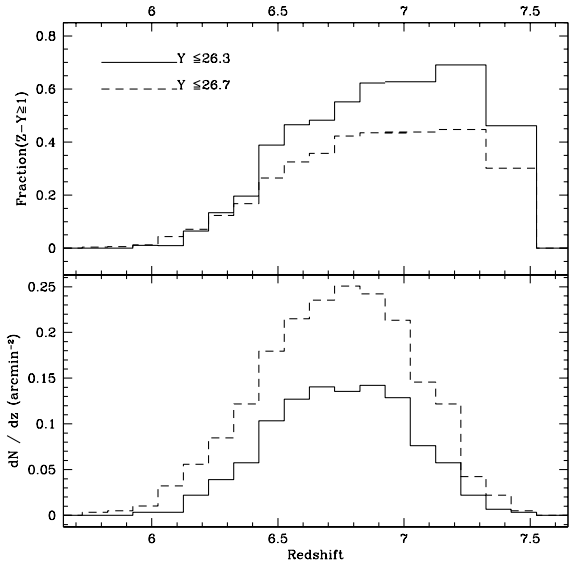}
   \caption{{\it Upper panel}: Fraction of $Y$-detected objects passing
     our criteria ($Z-Y\geq1$; non-detection blueward of $Z$) as a
     function of redshift, for two limiting magnitudes, $Y=26.3$
     (solid line) and $Y=26.7$ (dashed line). Input objects are
     extracted assuming a constant LF \citep{Mclure2009} from
     $z=6$. {\it Lower panel}: Redshift distribution (per unit
     redshift) for the same sample.}
         \label{simulnew}
\end{figure}

Finally, this set of simulations is used to estimate the systematic
effects acting when we use colours at shorter wavelengths, i.e. in the
$UBVRI$ bands. In these cases, because of the large IGM and internal
HI absorption, the expected flux in these bands for $z>6.5$ galaxies
is far below the detection threshold, or even null. For this reason a
stringent limit on the measured flux in these bands is adopted to
remove lower redshift interlopers.  However, the S/N estimated by
SExtractor may be a poor representation of the actual photometric
scatter at low fluxes, due to a combination of factors, such as
uncertainties in the estimate of the local background, underestimates
of the true r.m.s., or chance superposition of faint blue galaxies
along the line of sight.

To account for these effects, we have measured the resulting
signal--to--noise SN in the $UBVRI$ images for each simulated objects
inserted in the $Y$ one, which should be zero on average. It turns out
that the actual distribution of the S/N ratios is wider than the one
obtained with SExtractor, which is computed scaling the input weight
image.  With this set of simulations we thus estimate the
``effective'' r.m.s. $\sigma_{S/N}$, i.e. the r.m.s. of the
signal--to--noise distribution in each of the 5 images, which is
typically about 1.5.  Even taking into account this wider
distribution, we also find that the tails of the S/N distribution
($S/N>2 \sigma_{S/N}$) contain more objects than in the case of a pure
Gaussian distribution.

As mentioned above, we will use the estimated $\sigma_{S/N}$ in all
UBVRI bands, requesting that high-z candidates have flux
$<2\sigma_{S/N}$ in all UBVRI bands and $<1\sigma_{S/N}$ in at least
four of them.  With our simulation, we estimate that the fraction of
true high-$z$ galaxies lost because of this strict criterion is about
$\sim30\%$. This effect will also be taken into account in our
estimate of the LF.

The output of this exercise can be summarised in
Figure~\ref{simulnew}.  The final effect of noise on the $Z-Y>1$
colour is shown in the upper panel, where we plot the fraction of
simulated objects detected in $Y$ and with measured $Z-Y\geq1$ as a
function of redshift, for two limiting magnitudes, $Y=26.3$ (solid
line) and $Y=26.7$ (dashed line). We remind that, since the objects
are extracted from a steep LF, most of them are close to the limiting
magnitude.  The $Z-Y>1$ colour and the requested non-detection
blueward of $Z$ provide an effective cut below $z=6.3-6.4$, which is
nearly total at $z<6$, even including the effect of noise. At $z>7$,
the fraction of detected objects is less than 100\% because of the
combined effects of incompleteness in the detection and of the
requested non-detection blueward of $Z$, which rejects some genuine
high-z galaxies because of photometric scatter.

The most important output is shown in the lower panel of
Figure~\ref{simulnew}, where we plot the expected redshift
distribution of our sample, as extracted from the simulations. In this
case, counts are normalised assuming that the LF remains constant from
$z=6$ with the values provided by \citet{Mclure2009}. Since we
populate the input catalog following a LF with a realistic slope, we
expect that this redshift distribution is a good estimate of the
redshift selection function of our survey. The low redshift cut-off is
due to the $Z-Y>1$ colour and the requested non-detection blueward of
$Z$, as discussed above. At higher redshift, the cutoff is due to the
decreasing number of expected galaxies in a magnitude-selected sample
- as shown in Figure~\ref{DM}, objects of given $L$ become rapidly
fainter at $z>7.3$. In this case, the high redshift tails of the
distribution is sensitive to the evolution of the LF: in case of a
marked drop of luminous objects at high redshift, with respect to the
\citet{Mclure2009} LF, the distribution will be skewed toward the
lower redshift boundary.  This plot clearly shows the basic feature of
our $Y$-band selected approach, which allows to sample the LBG LF in a
well defined narrow redshift range. We note that this redshift range
is narrower than the corresponding one in $z-J$ selected samples .

\section{Detected z$>6.5$ Galaxies}
\label{7galaxies}
\begin{figure}
   \centering
   \includegraphics[width=8cm]{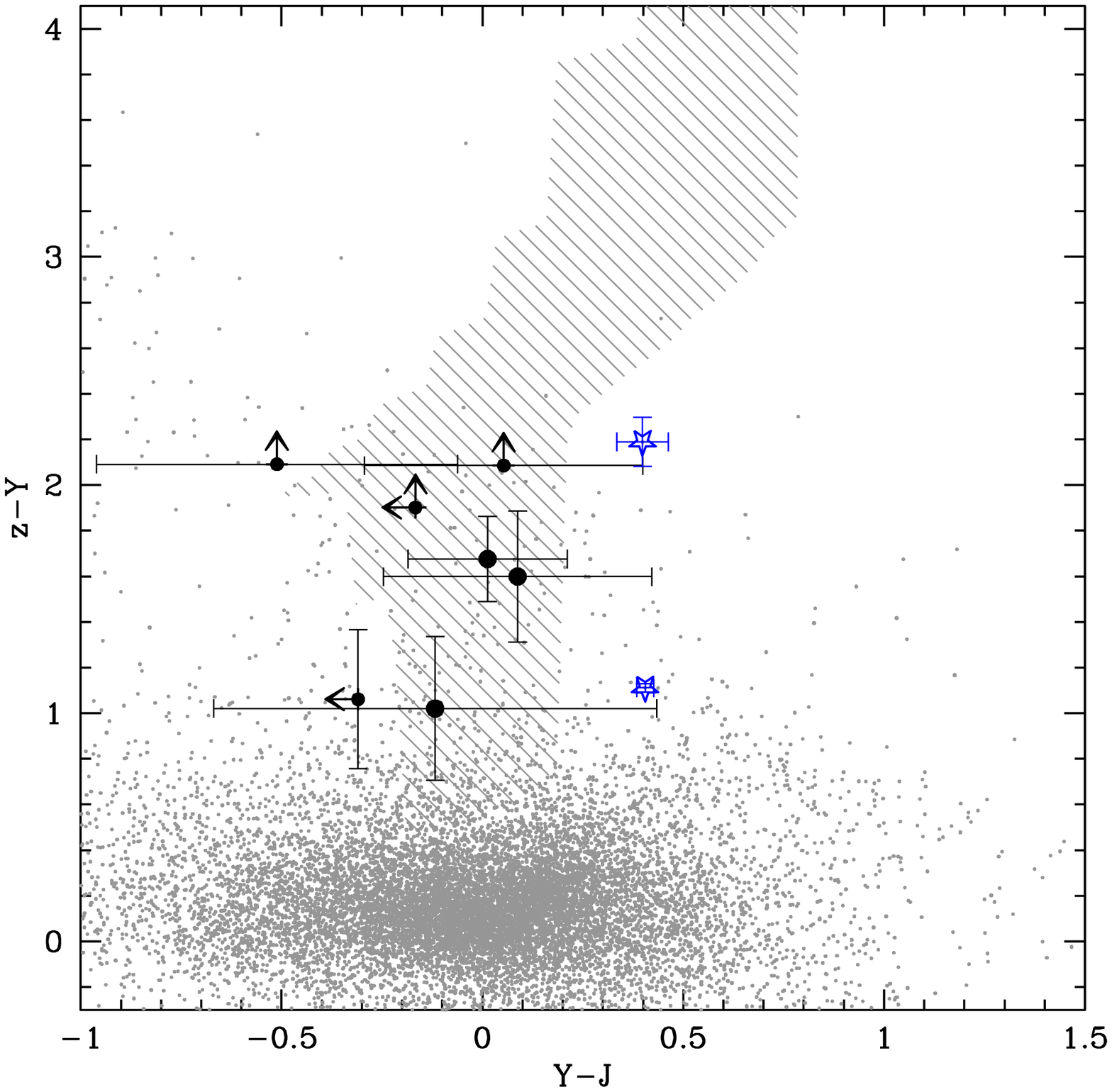}
   \caption{Position of the high-z candidates in the $Z-Y$ vs $Y-J$
     colour plane (black large dots). Upper limits are computed at the
     $1\sigma$ level. The shaded area shows the location of the
     expected colours of CB07 models, at $z\geq 6.5$. The faint gray
     dots show the rest of the sample. Objects fulfilling the
     selection criterion are rejected because of their detection in at
     least one of the UBVRI images (see text for details). The two
     Brown Dwarfs already identified by Mannucci et al. are shown as
     stars. The one in the upper right is selected as a high-z
     candidate on the basis of its UBVRIZY photometry, but removed
     from the final sample.  }
         \label{ZYJ}
\end{figure}

\begin{figure*}
   \centering
   \includegraphics[width=18cm]{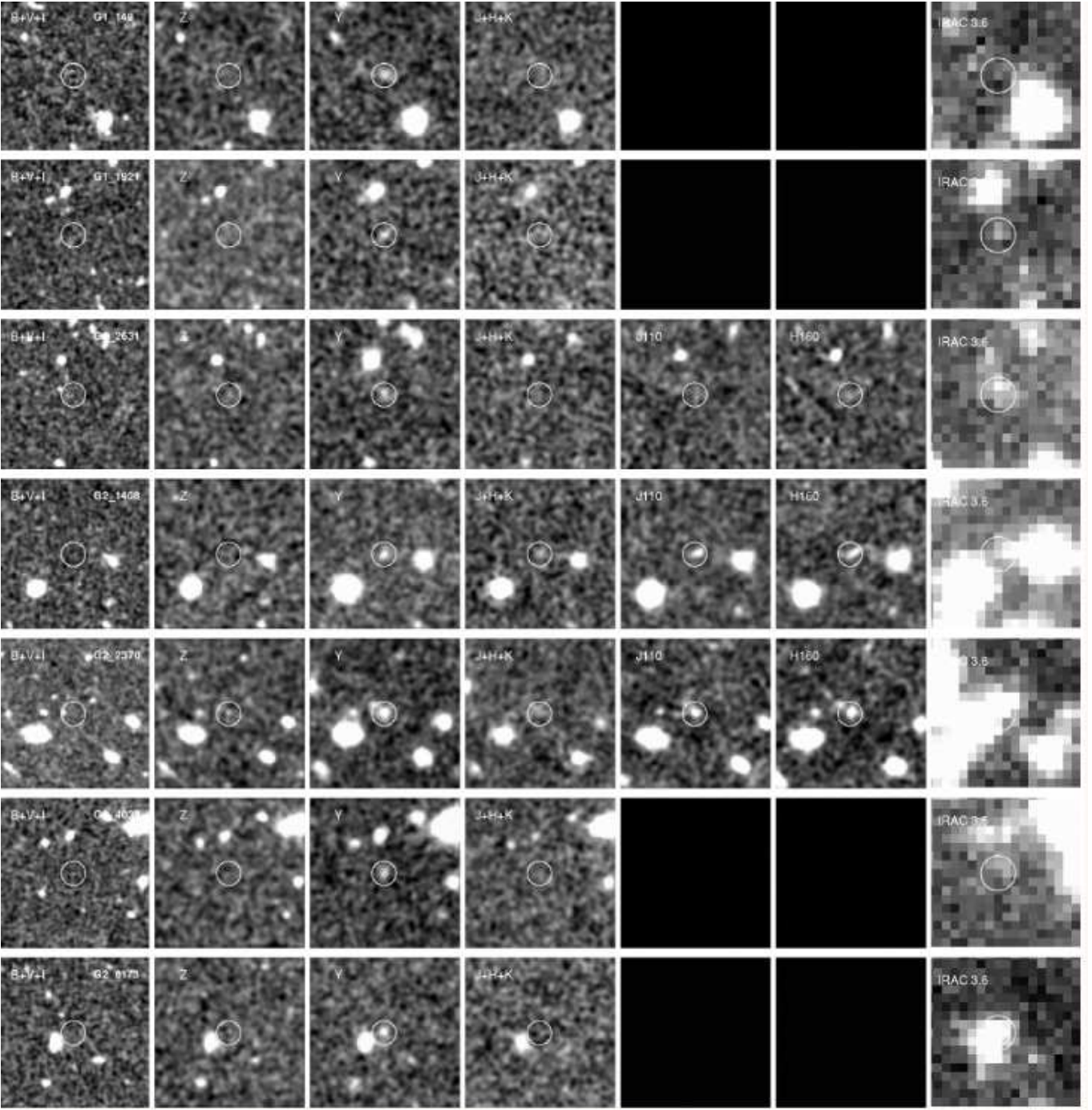}
   \caption{Thumbnails showing the images of the 7 selected high
     redshift candidates in the different observed bands.}
         \label{thumb}
\end{figure*}

Based on the results of the previous sections, we finally select the
$z\geq6.5$ candidates adopting the following criteria:

\noindent
1) We include only objects detected in the deepest regions of the
images (r.m.s. not larger than 1.5$\times$ its minimum value) with
total magnitude Y$\leq 26.7$; 
\\
2) We require:
\begin{eqnarray*}
(Z-Y) &>& 1.0 \\
(Z-Y) &>& 1.0+(Y-K)\\
(Z-Y) &>& 0.5+2.0(Y-J)\\ (Y-J)&<& 1.5 \\(Y-K)&<& 2.0
\end{eqnarray*}
\\

3) We require that the S/N (as measured in units of the
simulation-calibrated $\sigma_{S/N}$) is less than $2$ in all UBVRI
bands, and above $1$ in one of these bands only.

Figure~\ref{ZYJ} shows the position of all objects of our sample in
the $Y-J$ vs $Z-Y$ plane, with the final candidates and the two known
brown dwarfs shown highlighted.

In practice, the limit on the $Y-J< 1.5$ and $Y-K\leq 2$ colours are
ineffective. Indeed, all the objects with $Z-Y \geq 1$ are very blue
in the near-IR, having $Y-J< 1.2$ and $Y-K<1.7 $, and 90\% of them have
$Y-J< 0.5$ and $Y-K<1 $.

As a result, the critical requirement to select ``bona-fide'' high-z
candidates is the non-detection in all the bands blueward of the $Z$
filter. In our catalog we  detect 139 objects with $Z-Y \geq 1$
and $Y\leq26.8$, only 9 of which pass the additional criteria on the
non-detection in the $UBVRI$ bands.

Most of our candidates are marginally detected at very low $S/N\simeq
2$ in the $Z$ and/or $J$ images, which makes mandatory to critically
assess their reality.  The first and most obvious check to be done is
with the available NICMOS F110W and F160W images in the GOODS area.
Unfortunately, these images do not cover with the required depth the
whole Hawk-I pointings. Out of 4 of our candidates falling on suitably
deep images, we clearly detect 3 of them with F110W and F160W
magnitudes consistent with our $Y$ one. The fourth one is undetected
in the F160W with $m_{F160W}-m_{Y}>1$ (F110W is not available).  As a
consequence we remove this object from our sample.

A similar check can be done on the IRAC images. Taking into account
the actual depth of the IRAC images, a detection is not definitely
required to confirm the existence of these candidates. Indeed, using
the same synthetic libraries described above, we expect colours
$m_{Y}-m_{IRAC}$ in the range $-2 / +2$ mags, primarily depending on
the age, mass and dust content of the galaxies. Since the detection
limit in IRAC does not exceed 26.5, grossly the same AB limit of our
candidates, we can expect that only a fraction of these galaxies are
detected in the IRAC images. Indeed, 3 (out of 8) objects are clearly
detected in the IRAC $3.6\mu$ channel (IRAC36 hereafter): G1\_1921,
G2\_1713 and the object G1\_2631 already detected in NICMOS. 
Further three objects (G2\_1408, G2\_2370 and G2\_6173) are marginally
detected due to blending with nearby sources. We note that two of the
IRAC detected objects are the faintest ones in our sample (G1\_1921 and
G1\_2631), so the IRAC detection confirms their reality.

We computed aperture photometry (corrected to total) on both
 the 3.6$\mu$ and the 4.5$\mu$ IRAC image.  
 None of our IRAC detected objetcs shows a $(3.6-4.5)$ index larger
 than $\sim 0.4$, in agreement with the colours expected for high
 redshift LBGs. 
 On the other hand, we note that
  previous studies have found that cool dwarf stars 
 show a broad range in the $4.5\mu$ luminosity, yielding values in the
   $3.6-4.5$  colour index that can be much
 larger than those of our objects \cite[up to 2.0 magnitudes,
 see][]{Patten2006,Helling2008}, although lower colour 
 indexes have also been observed for these objects. 

As already described in Sect. 3.2, one of our candidates was already
detected as a potential high--z candidate by \citet{Mannucci2007} and
associated to a Galactic brown dwarf.

We show in Figure~\ref{ZYJ} the position of this and of the other
Galactic brown dwarf found by \citet{Mannucci2007}, as well as of the
other 7 candidates.  It is interesting to note that, given the high
S/N that we obtain on these relatively bright brown dwarfs, the
measured $Y-J$ colour is larger than that expected for a Lyman-break
galaxy at $z>6.5$. This analysis further supports the conclusions of
\citet{Mannucci2007}, therefore we remove G1\_1713 from our sample of
``bona-fide'' high-z candidates.

In the attempt to identify further brown dwarfs, we have also
investigated the morphology of the candidates in the NICMOS images,
where available. Object G2\_2370 has a full width of about $0.45"$,
close to the $0.4$ value in NICMOS, but with a non-null ellipticity
0.3. Object G2\_1408 is clearly resolved, with FWHM$\simeq
0.65$. Considering also that several LBGs with spectroscopic redshifts
in GOODS have a NICMOS PSF consistent with the stellar one, we shall
keep both of them in our sample.

After removing from our sample the \citet{Mannucci2007} brown dwarf
and the object undetected in deep NICMOS pointings, we are left with 7
candidates: their position and properties are given in Table 1.  The
relevant thumbnails are given in Figure ~\ref{thumb}.

\begin{figure}
   \centering
   \includegraphics[width=8.6cm]{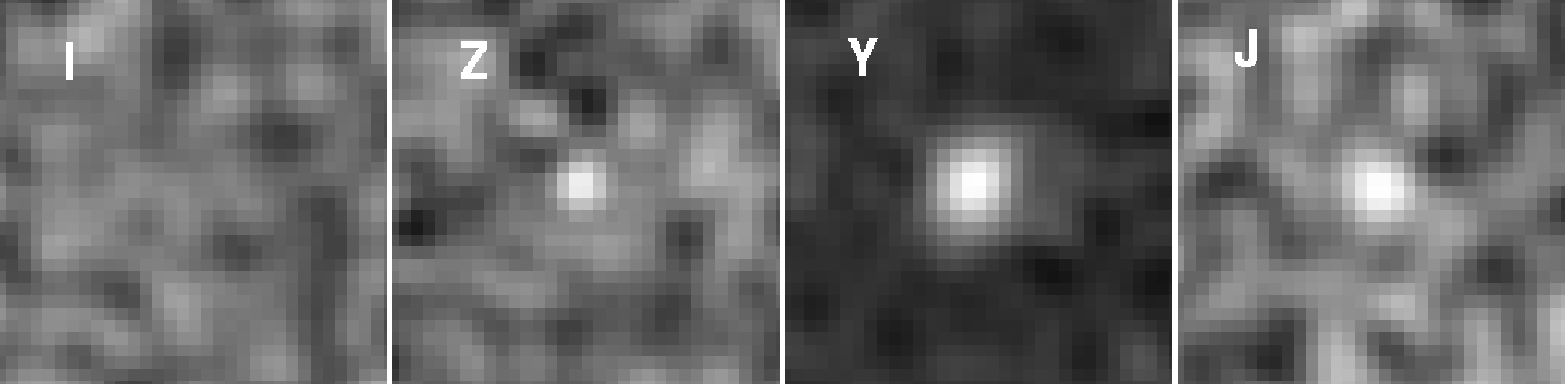}
   \includegraphics[width=8.6cm]{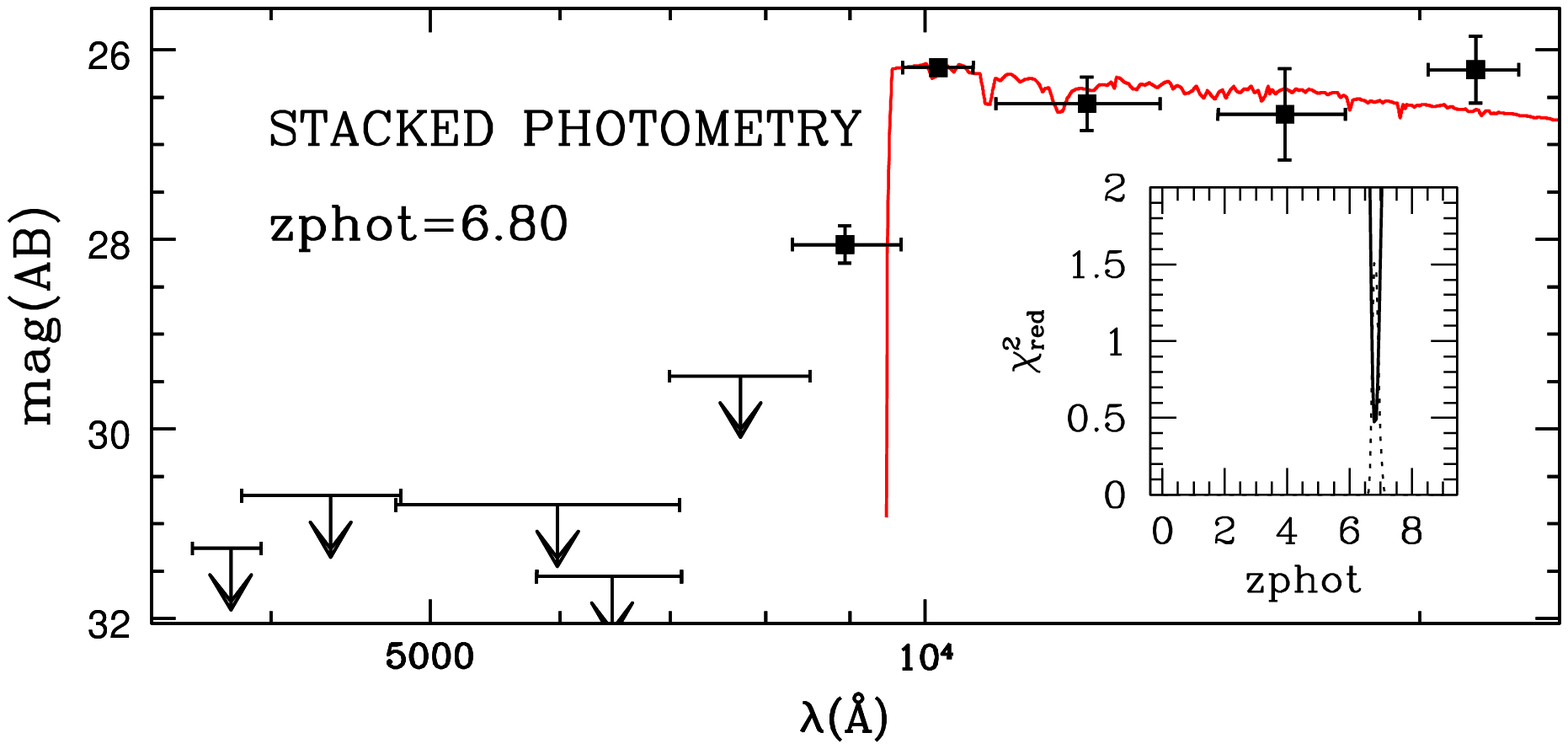}
   \caption{{\it Upper:}Thumbnails showing the stacked images of the 7
     selected high redshift candidates in the observed bands shown in
     the legends. {\it Lower}: resulting SED, with relevant
     photometric redshift at $z=6.8$}
         \label{thumb_stack}
\end{figure}

We performed a stacking of all the thumbnails in the UBVRIZYJHK images.
We confirm that objects are undetected in the UBVRI images, are
detected in the J band average image and exhibit an average colour
$Z-Y\simeq 2$. The resulting SED provides an excellent photometric
redshift at $z=6.8$. Relevant thumbnails and SED are shown in Figure~
\ref{thumb_stack}.

We checked if any of the previously published, NICMOS
detected, $z>6$ candidates in the UDF area
\citep{Bouwens2004,Bouwens2006b,Oesch2009} are present within our
sample.  We find that our candidate $G2\_1408$ is the object
UDF-387-1125 in \citet{Bouwens2004}. All the other objects are not
detected with our extraction parameters. We notice however that most
of these objects have faint magnitudes in the J110 and H160 NICMOS
bands ($\sim 26.8 - 27.0$ AB), while we adopted conservative selection
criteria in order to have a low number of spurious detections at the
faintest magnitudes in our sample ($Y = 26.8$, see
Sect. \ref{detection}), so it is not surprising that we are missing
objects detected in the very deep NICMOS pointings over the UDF.
We note that the colours of all the selected objects are
consistent with small values of $Y-J$ (or $Y-K$) plane, as shown in
Figure~\ref{ZYJ}. Since larger value of $Y-J$ (even within our
threshold $Y-J\leq 1.5$) are typical of galaxies at $z>7.2$, this
confirms our expectations that the redshift distribution is in
practice limited at this value.

We also checked other recent results on the GOODS area presented by
  \citet{Hickey2009} and by \citet{Wilkins2009}.
 
 \citet{Hickey2009} exploit
 part of the same Hawk-I data-set presented here but reach a shallower
 magnitude limit (Y$\sim$25.7-25.9).  Their candidate  ID 9697 is our candidate
 G2\_2370, while their object ID 9136 is rejected in our sample beacuse of
  S/N in the I band slightly above our threshold. Their other two candidates show significant
 detection in the bluer bands and so are probably lower redshift interlopers
 as discussed also by the authors.

  \citet{Wilkins2009} present an analysis of
 the WFC3 ERS data covering the northern portion of our ``GOODS2''
 field. Their candidate ~\#4 is our object G2\_6173, although they 
 measure a fainter Y magnitude. Of their two candidates having Y$<$26.8  we
 reject object ~\#1 beacuse of non negligible S/N in the I band, while object ~\#2
 is too close to a bright (Y$\sim$20) source to be effectively de-blended in
 our ground-based images and it is not present in our catalogue.
 
We remark that we estimated that a 30\% of real sources are
  missed with the strict rejection criteria in
 the  blue bands adopted in the present analysis. This is 
 consistent with object ID 9136 in \citet{Hickey2009}
  and object ~\#1 in \citet{Wilkins2009} being real high redshift galaxies
  whose high I-band S/N in our catalogue is due to photometric scatter,
  although their non negligible S/N in the I band might be due to a redshift
  near the lower limit of the redshift selection window. 
  We remark that potential missed detections beacuse of photometric scatter,
  or beacuse of objects falling near other bright sources, are fully taken
  into account in the LF estimate
  presented in the next section based on extensive imaging
  simulations.
\begin{table}
\caption{Hawk-I z-drop candidates in Goods-S}
\label{data}
\centering
\begin{tabular}{ccccc}
\hline
ID & R.A. (deg) & DEC. (deg)& Y & Z-Y \\
\hline
G1\_149& 53.157038 &-27.930707 & 26.06& 1.60\\
G1\_1921&53.19538 & -27.835421& 26.71&  1.02\\
G1\_2631& 53.106022 & -27.848123& 26.64&1.06 \\
G2\_1408& 53.177382&  -27.782416&26.37 & $>$2.1\\
G2\_2370& 53.094421& -27.716847&25.56 &1.68 \\
G2\_4034& 53.150019& -27.744914& 26.35& $>$2.1\\
G2\_6173&  53.123074& -27.701256& 26.53& $>$1.9\\

\hline
\end{tabular}
\end{table}

We finally checked whether there is any X--ray emission detectable in
the deep X ray images \citep{Luo2008}. No source was individually
detected. In the stacked image, the 90\% count rate (0.5-2keV) limit
is $2.41\times 10^{-6}$ counts/s, for a total exposure time of
$1.2\times 10^{7}$s. The resulting Luminosity is $\leq 7.8 \times
10^{42}$ ergs/s in the band from 2 to 10 keV (assuming a power--law
spectrum with a energy index -0.4). Assuming a bolometric correction
factor of 20 and a Eddington--limited accretion rate, we derive a
limit on the black hole mass of about $\leq 10^{6}M_{\odot}$.

\section{The Evolution of the LF}
\label{LF}

To estimate from our detections the most likely LF
we use an approach that fully accounts for the expected systematics in
the detection process, as done by, e.g., \citet{Bouwens2007,Mannucci2007,Mclure2009}. We refer in the following
to the luminosity $L$ as measured at $1500$\AA ~rest-frame.

First, we assume that the LF can be described by the usual Schechter
function with parameters $\phi$, $\alpha$ and $M_*$ \citep{Schechter1976}.  Unfortunately,
we are unable to constrain the slope of the LF, since our faint limit
is close to the expected value of the characteristic luminosity
$M_*$. For this reason, our results are grossly insensitive to the
value of the slope index $\alpha$, that we shall fix to the value
$\alpha=-1.71$ of the $z \sim 6$ LF by \citet{Mclure2009}, close to
$\alpha=-1.74$, used by \citet{Bouwens2007}. We have explicitly tested this
assumption by fixing $\alpha$ to significantly different values
($\alpha = -1.4, -2.0$), without finding significant differences.

We will assume that the other two parameters evolve in
redshift from their value at lower $z=z_0$ value:
$$log(\phi(z)) = log(\phi(z_0)) + dlog(\phi)/dz \cdot (z-z_0)$$
$$M_*(z) = M_*(z_0) + M_*' \cdot (z-z_0)$$

\begin{figure}
   \centering
   \includegraphics[width=8cm]{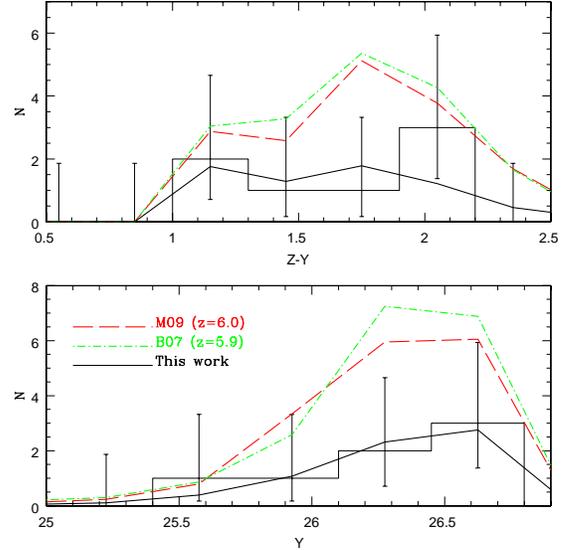}
   \caption{$Y$ and $Z-Y$ distributions of the high redshift candidates and of our best-fit evolving LF at $z>6$ (black solid line), compared with  the
     expected values from recent estimates of the LF at $z\sim 6$ by \citet{Bouwens2007} (B07, green dot-dashed line) and \citet{Mclure2009} (M09, red dashed line), as computed according to the simulations described in Sect. \ref{LF}. }
         \label{nz}
\end{figure}

We note that our parametrisation is analogous to the one adopted by
\citet{Bouwens2008}. While \citet{Bouwens2008} used this
parametrisation to fit their values at different redshifts (in the
range $4<z<10$), we explicitly vary the Schechter parameters within
our redshift range $6.4<z<7.2$, which might be important in case of a
strong evolution. We have also tested a purely linear evolution of
$\phi$, which turned out to produce consistent results.

In principle, all four parameters could be left free in the
minimisation process.  However, given the size and depth of our
sample, we shall assume for $ M_*(z_0)$ and $ \phi(z_0)$ the observed
values at slighter lower redshifts $z\simeq 6$ and evaluate the
evolutionary terms $ M_*'$ and $ dlog(\phi)/dz$ alone. 

During the minimisation process, for any given value of the free
parameters, we Monte Carlo extract a sample of objects with redshift
$z$, 1500\AA ~rest-frame absolute magnitude $M$, randomly adding
different E(B-V) and metallicities as obtained by the CB07 models
shown in Figure~\ref{DM} and Figure~\ref{Z_Y}. We include Ly$\alpha$
emission with a Gaussian distribution with standard deviation
30\AA. All these galaxies are extracted from the (larger) simulations
described in Sect.~\ref{simulations}. This way, we can convert their rest-frame
$M_{UV}$ into observer-frame magnitudes in the same bands used to in
our observed sample, taking into account all the uncertainties
involved in the observations discussed in Sect.~\ref{simulations}:
detection completeness, photometric scatter and random fluctuations in
the S/N measure due to overlapping interlopers or other effects.

\begin{figure}
   \centering
   \includegraphics[width=9cm]{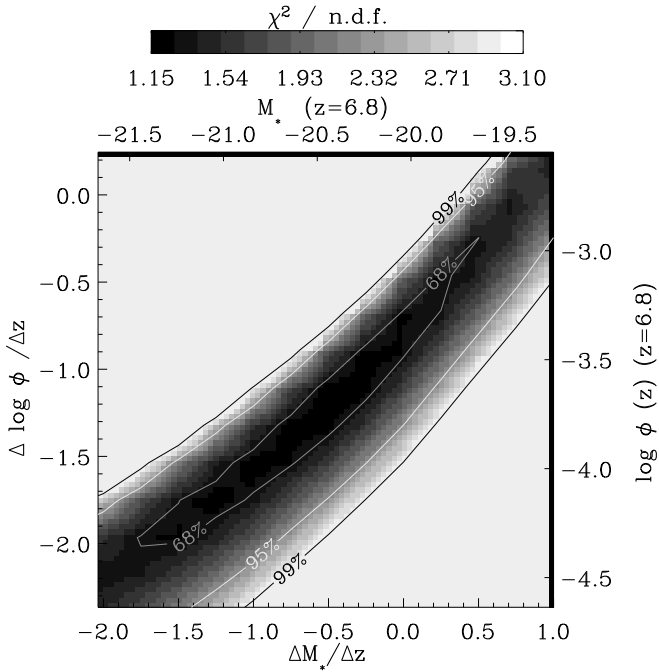}
   \caption{$\chi^2$ contour levels for the $ dlog(\phi)/dz$, $ M_*'$
     parameters derived for the Schechter--like LF.  The lower and left axis
     refer to the evolutionary terms $ M_*'$ and $ dlog(\phi)/dz$ (see text for details), the
     upper and right axis refer to the  $ M_*$ and $ \phi$ values at the median redshift estimated for our sample (z=6.8). }
         \label{chi2}
\end{figure}
\begin{figure}
   \centering
   \includegraphics[width=10cm]{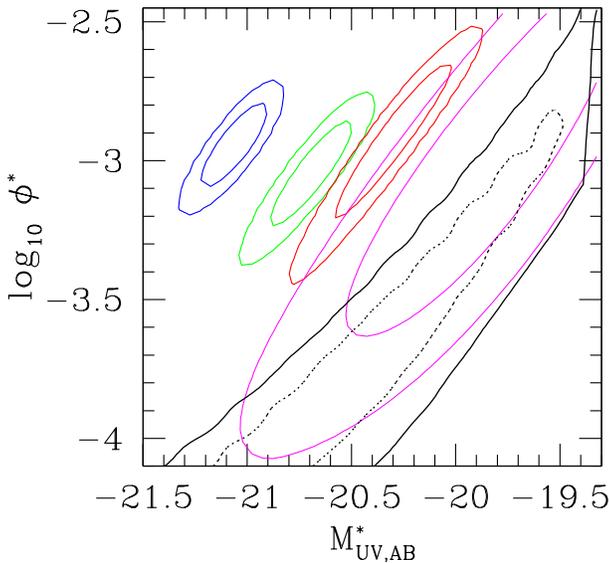}
   \caption{$68\%$ (dotted line) and $95\%$ (continuous line) $\chi^2$
     contour levels for the value of $M_*$ and $ \phi$ at $z=6.8$ as
     estimated from our sample of Y-detected LBGs. Blue, green, red
     and magenta contours are the 68\% and 95\% likelihood intervals
     on the Schechter parameters as estimated by \citet{Bouwens2007}
     at redshifts 3.8, 5.0, 5.9 and 7.4 respectively. }
         \label{chi2_bow}
\end{figure}

The distributions of magnitudes and colours for each Monte Carlo
simulation are scaled to the observed area in the GOODS-S field (after
excluding the fraction of area lost because of the presence of lower
redshift objects) and compared to the observed ones with a maximum
likelihood test under the assumption of simple poissonian statistics.

To provide a visual example of the significance of the results, we
show in Figure~\ref{nz} the expected number counts for three cases of
the evolution of the LF, comparing them with our observed counts.
It is immediately appreciated that, barring for the moment large
fluctuations due to cosmic variance, the number counts of our
candidates imply a strong evolution of the LF in the relatively short
cosmic time elapsed from $z\simeq 6$ to $z=6.8$.

To constrain the evolution of the LF, for each simulated population,
and for each of the two distributions, we build the likelihood
function $\cal{L}$
\begin{equation}
 {\cal L} = \prod_{i} e^{-N_{exp,i}} \frac{(N_{exp,i})^{N_{obs,i}}}{(N_{obs,i})!}
\end{equation}
where $N_{obs,i}$ is the observed number of sources in the magnitude
(colour) interval $i$, $N_{exp,i}$ is the expected number of sources
in the same magnitude (colour) interval, and $\Pi_{i}$ is the product
symbol. The final likelihood assigned to each given model is the
product of the likelihoods computed for the magnitude and colour
distributions separately.  We iterate this procedure for a grid of ($
dlog(\phi)/dz$, $ M_*'$) values.  

The colour plot in Fig. \ref{chi2} shows the 68\%, 95\% and 99\%
likelihood intervals on the evolutionary terms $ M_*'$ and $
dlog(\phi)/dz$ (left and bottom axes) and for the resulting Schechter
parameters at the median redshift z=6.8 of our sample (top and right
axes). In the same plot, the colour code refers to the $\chi^2$
distribution obtained under the usual assumption $\chi^2=-2.0\cdot
ln(\cal{L}) $.  It is evident that the absence of any evolution in
both parameters ($dlog(\phi)/dz=0$ and $M_*'=dM_*/dz=0$)
with respect to the best-fit values at $z=6$ is ruled out at $\gtrsim$
99\% confidence level. This is shown more clearly by
Fig. \ref{chi2_bow} where our 68\% and 95\% likelihood intervals at
z=6.8 are displayed together with the likelihood contours computed in
a self-consistent way by \citet{Bouwens2007} at redshifts 3.8, 5.0,
5.9 and 7.4.

The formal maximum of the maximum likelihood lays at $d log(\phi) /
dz = -0.89\pm0.21$ and $M'=dM_*/dz= -0.25\pm0.56$ (errors include the
effect of cosmic variance, see below). At face value, our best fit
model indicates mostly a decrease in the normalisation factor $\phi$
when compared to the best fit parameters at $z\sim 6$
(Fig. \ref{chi2_bow}). 
However, the relatively small number of galaxies that
we have in our fields does not allow us to put tight constraints on
the exact combination of Schechter parameters: as shown by the
elongated shape of the contour levels in Fig. \ref{chi2_bow} the two
parameters are highly degenerate. We indeed find that the position of
the maximum is poorly constrained, and somewhat dependent on the
details of the simulations, like step or number of simulated
galaxies. We therefore provide its value only for reference, and will
focus on the overall range of allowed values at the 1$\sigma$ contour
or on integrated quantites. 

The effects of cosmic variance are a significant concern in our case,
since all our data come from a single pointing which includes most of
the previous NICMOS--based surveys. Indeed, cosmic variance effects
are not independent from those potentially affecting the
\citet{Bouwens2008} results.  The cleanest way to get rid of these
effects is the analysis of several independent fields applying the
same selection criteria.  For this reason, our survey will continue by
covering two completely disjoint fields with the same setup. For the
moment, we can evaluate the possible impact of cosmic variance by
measuring the relative variance within 200 samples bootstrapped from
the Millennium Simulation presented by
\citet{kitzbichler2007}, that is considered to accurately reproduce 
clustering properties even at very high redshift \cite[e.g.][]{Overzier2009}. We use an area as
large as our Hawk-I data set and we apply a corresponding photometric selection
criteria on galaxies at $6.4<z<7.2$, without any constraint on the
distribution of host halos.
 We estimate that a cosmic
variance of $\sim 30\%$ affects the number counts of z-drop LBGs in
our $\sim 90 arcmin^2$ area. We have estimated the resulting effect on
the $\phi$ and $M_*$ parameters changing by $\pm 30\%$ the observed
number counts, and finding the relevant best-fit parameters. These
offsets were added in quadrature to the Hessian errors computed around
the best fit value.

In this way we obtain a best fit interval for the evolutionary terms
along with their associated uncertainty which includes both Poisson
and cosmic variance noise: $d log(\phi) / dz = -0.59\pm
0.25$ and $M'=dM_*/dz= 0.13 \pm 0.51$, resulting in
$\Phi^*=0.35^{+0.16}_{-0.11} 10^{-3} Mpc^{-3}$ and
$M^*(1500\AA)=-20.24\pm0.45$ at z=6.8.

Fortunately, the degeneracy in the $M_*$ and $\phi$ best-fit values is
not reflected in a comparable uncertainty in the number density of
bright galaxies, i.e. in the integral of the LF up to $M_*$. Indeed,
because of the correlation between the acceptable values of $M_*$ and
$\phi$, an increase in the normalisation ($\phi$) is compensated by a
decrease in the characteristic luminosity, such that the integral up
to $M\simeq M_*$ remains grossly constant. Integrating the best-fit UV
LF up to $M=-19.0$ we obtain an UV luminosity density $\rho_{UV}=
1.5^{+2.0}_{-0.9} 10^{25} erg ~ s^{-1} ~ Hz^{-1} ~ Mpc^{-3} $. The
errors are computed integrating all the UV LF that are acceptable at
the 68\% level.

For comparison, the integral of the z=6 UV LF of \citet{Mclure2009} up
to the same magnitude limit yields a $\rho_{UV}= 5.6^{+3.1}_{-2.3}
10^{25} erg ~ s^{-1}$. Our estimate thus implies a drop of a factor
$\sim 3.5$ in the UV luminosity density from z=6 to z=6.8.

We can convert this value in a star formation rate density following
the standard formula by \citet{Madau1998} and applying the extinction
correction of \citet{Meurer1999} (considering an average UV slope
$\beta=-2.0$). We obtain $SFRD=2.8^{+2.7}_{-1.6} 10^{-3} M_{sun} ~
yr^{-1} ~ Mpc^{-3} $.

\begin{figure}[!h]
   \centering
   \includegraphics[width=8cm]{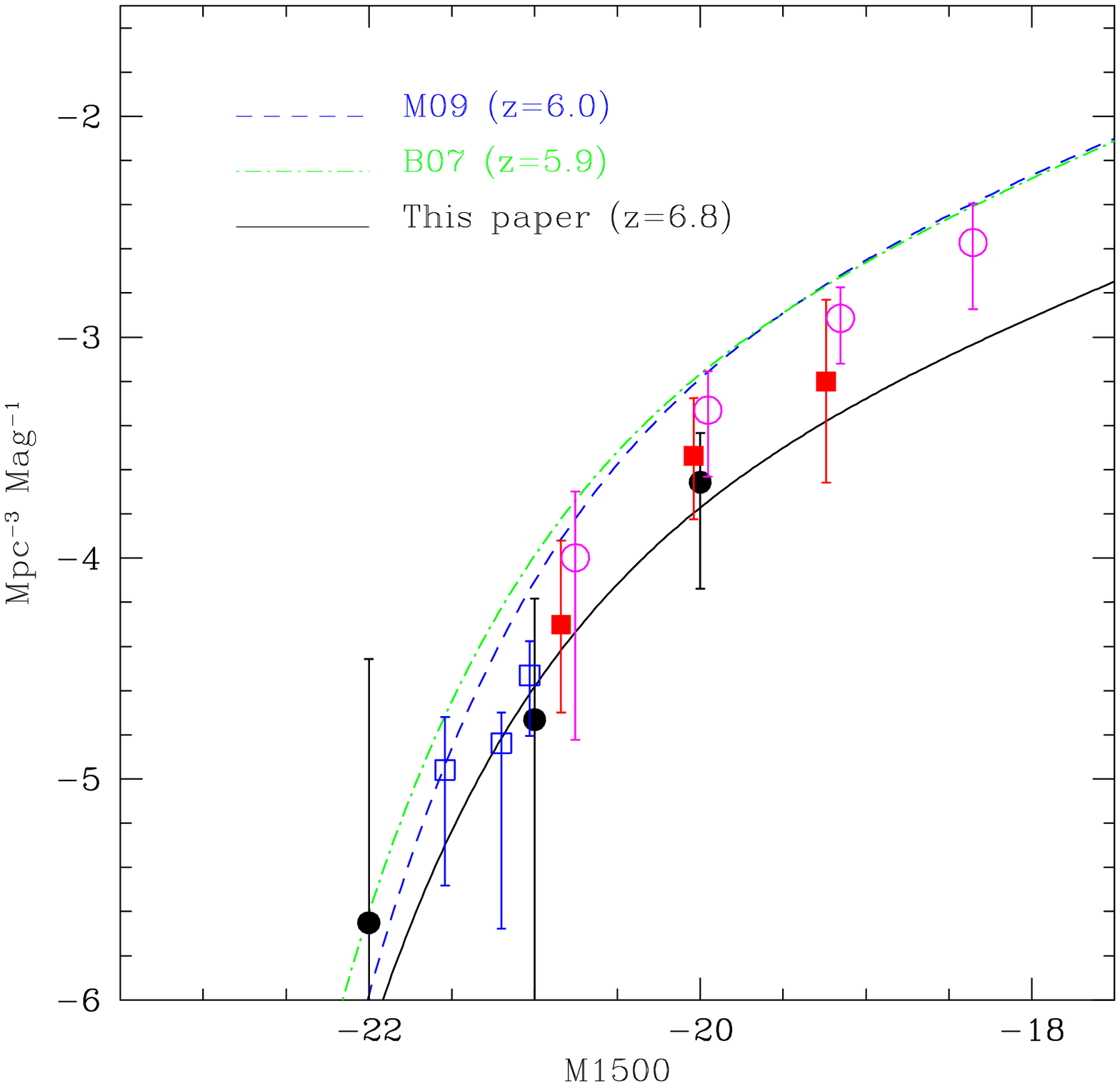}
   \caption{Number densities in three rest-frame magnitude intervals estimated
     for our Hawk-I data-set in a stepwise form (black circles and error
     bars), along with the results by \citet{Bouwens2008} for NICMOS
     detected objects (red filled squares),
     \citet{Ouchi2009} (SUBARU, blue empty squares) and \citet{Oesch2009b}
     (WFC3, magenta empty circles). The black solid line is our best-fit LF
     discussed in the text. For a comparison we show the recent determinations
     of the LF at $z\sim 6$ by \citet{Bouwens2007} (B07, green dot-dashed line) and \citet{Mclure2009} (M09, blue dashed line).}
         \label{stepwise}
\end{figure}
\begin{table}
\caption{Stepwise determination of the UV LF at z$\sim$6.8}
\label{Tab_stepwise}
\centering
\begin{tabular}{cc}
\hline
$M_{1500}$ & $\phi ~ (10^{-4} Mpc^{-3}~ mag^{-1})$\\
\hline
-22.0& $0.02 ^{+0.33} _{-0.02}$\\
-21.0& $0.19 ^{+0.47} _{-0.19}$\\
-20.0& $2.20 ^{+1.47} _{-1.47}$ \\
\hline
\end{tabular}
\end{table}

To provide a more straightforward comparison to other results
  in the literature, we have also computed the luminosity function for
  our z-dropout sample in a stepwise form \cite[see, e.g.][for a
  discussion of this method]{Bouwens2008}. Briefly, the stepwise
  method assumes that the rest-frame luminosity function of galaxies
  can be approximated by a binned distribution, where the number
  density $\phi_i$ in each bin is a free parameter: this non-parametric
  approach allows us to constrain the number density of galaxies at
  different magnitudes without assuming a Schechter-like shape.

We have assumed that the LF is made of three bins in the
  interval $-22.5<M_{1500}<-19.5$, corresponding to the range sampled
  by our observations. We also assume that galaxies are uniformly
  distributed within the bins, with number densities $\phi{_i}$ to be
  determined. We exploit the same set of simulations described in 
  Sect.~\ref{simulations} to compute the distribution of observed magnitudes
  originated from each bin, scaled to the observed area in the GOODS-S
  field. We then find the combination of $\phi{_i}$ that best reproduces
  the magnitude distribution of our observed objects with a simple
  $\chi^2$ minimization.  

The results are reported in Table~\ref{Tab_stepwise} and are
  displayed in Fig.~\ref{stepwise} along with recent results from the
  literature and our best-fit Schechter LF discussed above. 

Our two independent determinations of the LF at z$\sim 6.8$ (stepwise and
maximum likelihood) are in perfect agreement. The agreement at the bright end with the  densities estimated by
\citet{Bouwens2008} and \citet{Ouchi2009} is remarkable, considering the
widely different data-sets and selection techniques used and the large
uncertainties given by the low number of objects in our sample. The
disagreement at the faint end between our best-fit LF and the points by
\citet{Oesch2009b} is probably
 the result of the fixed slope $\alpha=-1.71$ we had
to assume for our maximum likelihood test, as discussed above, and it is
consistent with a steepening of the faint end, as suggested by the authors.

\section{Predictions on future surveys}
\begin{figure}
   \centering
   \includegraphics[width=8cm]{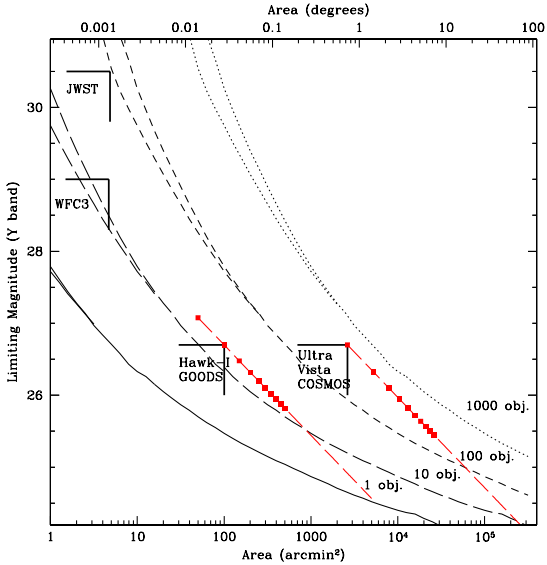}
   \caption{Expected number of $6.4<z<7.2$ galaxies as a function of
     area and limiting magnitude for present and future ground--based surveys
     and for deep pointings  with HST-WFC3 and
       JWST-NIRCam (see text for details).  
     Black curves show the combination of area and limiting
     magnitudes necessary to collect 1, 10, 100 or 1000 galaxies in
     this redshift range. Estimates are based on our best-fit values of $ dlog(\phi)/dz$,
     $ M_*'$ and considering the two values $\alpha$=-1.7 and $\alpha$=-1.9 (upper
       and lower curves, respectively). 
     Note that incompleteness and noise effects
     are not included in the computation and could significantly
     decrease the number of candidates eventually detected. The red
     straight lines show the  position of the two surveys if
     they were conducted over more pointings with the same total
     exposure time. Each dot represent an additional pointing.  }
         \label{survey1}
\end{figure}

It is interesting to use the values of the Schechter parameters found
in the previous section to estimate the number of detections that may
be obtained with future surveys. Clearly, the large errors in our
determination make this exercise uncertain, but nevertheless useful to
design future surveys from ground and space.

From the ground, the relative efficiency of the $Y$ band allows us to
cover significantly large areas of the sky.  This is useful both to
beat cosmic variance as well as to identify potential spectroscopic
targets. Using our best fit values of $ dlog(\phi)/dz$, $ L_*'$ (and
considering the two values $\alpha$=-1.7 and $\alpha$=-1.9) in Figure~\ref{survey1} we show the combination of
area and limiting magnitude necessary to detect a given number of
galaxies at $6.4<z<7.2$. We also show the position of our present data
(labelled as Hawk-I GOODS), of the planned Ultra-VISTA observations
of the COSMOS field, of the WFC3 pointing over the UDF \citep{Oesch2009b}
  and of a  $\sim$20hr pointing with the 115W filter of JWST-NIRCam. 
For the Ultra-VISTA COSMOS survey, we consider the case of the deepest
non-contiguous observations for a total of 0.73 deg$^2$.  We remark
that these predictions do not include the fraction of galaxies lost
because of different effects. In our case, the major sources of losses
are the incompleteness in the photometric detection and the scatter
(intrinsic and observational) in the $Z-Y$ colour that we use as
threshold, leading to a loss of about 30\% of the candidates. For this
reason, the number of candidates that we detect is about half of the
expected numbers in Figure~\ref{survey1}. Clearly, surveys with
similar incompleteness levels should expect a comparable reduction of
the observed numbers.  The regression of the bright side of the LF,
combined with its exponential slope, results in a flattening of the
expected cumulative numbers at bright fluxes even in wide areas.  In
practice, it is very difficult to detect galaxies brighter than 25 mag
even over areas of about 1deg$^2$, such that the detection of galaxies
brighter than 24.5 in Ultra-VISTA would be a formal violation of our
LF.

The two red straight lines in Figure~\ref{survey1}, show the expected
position in the area-limiting magnitude plane of the two ground-based surveys
(Hawk-I GOODS and Ultra-Vista COSMOS) if they were conducted over more
pointings with the same total exposure time (32hrs and 150hrs,
respectively). Increasing the number of pointings the total number of
candidates decreases, with a loss of statistical robustness.  For a
relatively small number of pointings, this is likely counterbalanced
by the decrease in the scatter due to cosmic variance. In essence, for
a given instrument and total investment of time, the highest return in
a statistical sense is given by a modest number of independent
pointings.

\begin{figure}
   \centering
   \includegraphics[width=9cm]{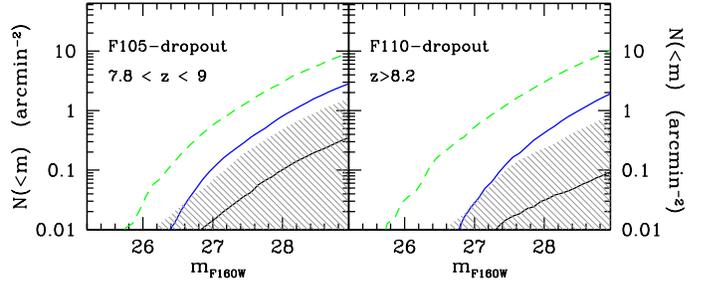}
   \caption{Expected cumulative number counts of high redshift
     galaxies as a function of the $F160W$ 
     magnitude, for different redshift ranges and different choices of
     the evolutionary LF. The adopted colour cuts and the
     corresponding redshift ranges are shown in the legend. In both
     panels. the green dashed line is the prediction from a constant
     (from $z=6$ to $z=10$) UV LF. The black solid line is our best
     fit evolving LF, while the shaded area shows the predicted number
     counts for all the LF parameters acceptable at the $1\sigma$
     level in our survey. The blue continuous line is the evolving LF
     from \citet{Bouwens2008}.  }
         \label{survey2}
\end{figure}

Only WFC3 on HST  has the required IR sensitivity to detect
galaxies at even higher redshifts, ahead of the advent of JWST. For
this reason, several surveys are starting or are being planned using the WFC3 IR
filters.  In Figure~ \ref{survey2} we provide the predicted cumulative
number counts for galaxies in different redshift ranges, selected
using either $F105M-F125W\geq 1$ (that corresponds to the redshift
range $z=7.8-9$) or $F110W-F160W\geq 1$ ($z\geq 8.8$). In both panels,
the magnitudes in the $F125W$ and $F160W$ filters are computed from
$L_{1500}$ using the same CB07 described above. We provide the number
counts expected in the case of a non-evolving LF from $z=6$ (clearly
an upper limit), the predictions from the evolving LF of
\citet{Bouwens2008} (which was drawn from $z=4$ to $z=10$) and the
extension to $z=10$ of our LF. For the extrapolation of our LF we also
show the whole region predicted extending from $z=7$ to $z=10$ all the
LFs acceptable at the $1\sigma$ level found from our analysis at
$z=7$.

Apart from the unrealistic case of a constant LF beyond $z=6$, the
number density of LBGs at $z\geq 8$ is expected to be very low, and
requires a combination of large areas and extreme depth to collect a
sizeable sample. Taking into account the WFC3 size (about 4.6
arcmin$^2$), it is necessary to reach continuum magnitudes at least
$m_{F125W}\simeq m_{F160W}\simeq 28$ to detect at least one dropout in
a single WFC3 pointing.

Figure~ \ref{survey2} also shows that the evolution of the LF resulting
from this work is faster than the \citet{Bouwens2008} one. Assuming
that this evolution continues at the same rate at higher redshift, the
number of expected dropouts would be smaller at higher
redshifts. Clearly, deep and wide observations with WFC3 will be able
to disentangle the various scenarios for the evolution of the UV LF.

\section{Summary and Conclusions}

We present in this work the results of a $Y$--band survey of the
GOODS-South field, aimed at detecting galaxies at $\gtrsim 6.5$ and
measuring their number density.

With this purpose we have made use of the deep $Y$--band observations
of the GOODS-South field obtained with Hawk-I, the new near-IR camera
installed at the VLT. We have matched and combined these data with the
publicly available images in the BVIZ (ACS), U and R (VIMOS) and JHK
(ISAAC) bands. The final area covered by these observations is of
about 90 arcmin$^2$ at a magnitude limit $Y\simeq 26.7$.

We analyze this sample to select high-redshift ($z\gtrsim6$) candidate
galaxies following a Lyman-break colour criterion adapted to our
filter set.  Galaxies are selected in the $Y$ band, and identified by
their large colour break $Z-Y>1$.  Particular care has been taken in
removing interlopers of various origins, including lower redshift
galaxies with large Balmer breaks, variable sources and Galactic brown
dwarfs, although some residual contamination from the latter cannot be
excluded.  An additional class of interlopers, exhibiting large $Z-Y$
colours as well as significant emission in the blue bands has been
removed by requiring a stringent non-detection in the UBVRI bands. We
argue that some of these contaminants might be very faint
emission-line galaxies or AGNs at intermediate redshift. The accuracy
and statistical impact of these criteria have been evaluated with
extensive Monte Carlo imaging simulations.

From the same simulations we estimate that our redshift
selection function is mostly efficient in the interval $6.4 \lesssim z
\lesssim 7.2$. 
   
We eventually isolated 7 highly reliable z-drop candidates after
removing from the colour selected sample one known galactic cool dwarf
star and one source undetected in deep available NICMOS F160W images.

To estimate the constraints that our observations set on the evolution
of the UV Luminosity Function at $z>6.5$ we ran detailed and realistic
imaging simulations of galaxy populations following different UV
Schechter functions with linearly evolving parameters $log(\phi)$ and
$M_*$ and a fixed value $\alpha=-1.71$. Our simulations account for
all the uncertainties involved in the observations: detection
completeness, photometric scatter and random fluctuations in the S/N
measure due to overlapping unresolved sources or other effects.  We
compare the resulting distributions of simulated magnitudes and
colours with the observed ones following a maximum likelihood approach
to constrain the parameters of the evolving UV LF.

We find strong evidence for an evolution of the luminosity function
above z=6: our analysis rules out at a $\sim 99\%$ confidence level
that the LF remains constant in both $\phi$ and $M_*$ above $z=6$,
even considering the effect of cosmic variance.

From our maximum likelihood analysis we estimate
$\Phi^*=0.35^{+0.16}_{-0.11} 10^{-3} Mpc^{-3}$ and
$M^*(1500\AA)=-20.24\pm0.45$ at the median redshift of our sample
($z=6.8$). With respect to the values found at $z=6$ our best fit
model indicates an evolution both in the normalisation factor $\phi$
and in the characteristic magnitude $M_*$. Our results are consistent
with the recent analysis by \citet{Ouchi2009}, considering the large
statistical errors and the high degeneracy between the two parameters.

Fortunately, the uncertainty and the degeneracy in the $M_*$ and
$\phi$ best-fit values are not reflected in a comparable uncertainty
in the number density of bright galaxies, i.e. in the integral of the
LF up to $M_*$.  We derive a UV luminosity density $\rho_{UV}=
1.5^{+2.0}_{-0.9} 10^{25} erg ~ s^{-1} ~ Hz^{-1} ~ Mpc^{-3} $
($M<-19$) and a star formation rate density $SFRD=2.8^{+2.7}_{-1.6}
10^{-3} M_{sun} ~yr^{-1} ~ Mpc^{-3} $. These values are definitely
lower than the corresponding ones at $z\sim 6$ by a factor $\sim 3.5$.

Although we did our best to carefully evaluate the systematic effects,
we caution that our results depend on an extensive set of simulations
to address the competing systematic effects which may increase or
decrease the observed number of candidates. We note that the two most
obvious - the possible residual contamination due to brown dwarfs and
the possibility of a spurious detection for two of our candidates -
will increase the amount of evolution from $z=6$ to $z=6.8$. Clearly,
only spectroscopic surveys and/or deeper imaging in the IR will
definitely settle the issue.

Such a strong evolution in the UV LF has important
consequences for reionization scenarios, as well as for the planning
of future surveys aimed at detecting very high redshift galaxies and
reionization sources.

Determining whether the UV emission of normal galaxies is capable of
reionizing the Universe at z$>6$ requires the knowledge of the value
of many parameters that are still unconstrained at these redshifts:
the escape fraction of ionising photons, the HII clumping factor, the
exact spectrum of star forming galaxies, their dust content as well as
the shape of the stellar initial mass function and the metallicity of
stellar populations
\cite[e.g.][]{Madau1999,Barkana2001,Stiavelli2004}. Following
\citet{Bolton2007} and their conclusions on the early estimates from
\citet{Bouwens2005} we note that the strong decrease we observe in the
UV emission coming from relatively bright sources implies that this
population alone is not capable of reionizing the Universe beyond
redshift 6. These observations can be reconciled with a completion of
reionization before z = 6 only under the hypothesis of an evolution of
the physical parameters quoted above, like an increase of the escape
fraction, an harder UV spectrum, a lower clumpiness factor or lower
metallicities \cite[e.g.][]{Henry2009,Oesch2009}. Another possibility
is that a relevant contribute to the UV emission comes from galaxies
at the faint end of the LF \citep{Bouwens2007} or from more exotic
sources \cite[see e.g.][]{Venkatesan2003,Madau2004}.

Thus it will be possible to fully assess the role of LBGs in the
reionization of the neutral IGM only putting tighter constraints both
on the bright and on the faint end of the luminosity function, at
z$\sim7$ and beyond.

Our results have also implications on the efficiency of these future
surveys. We show that to collect sizeable samples of $z>7$ bright
galaxies it is necessary to reach very faint limits ($m_{F160W}\gtrsim
28$) over relatively large areas. On the other hand, the extrapolation of our results to higher
redshifts indicates that dedicated surveys with HST-WFC3
will be able to disentangle the various scenarios for the evolution
of the UV LF parameters above $z\sim6$.

\begin{acknowledgements}

  Observations have been carried out using the Very Large Telescope at
  the ESO Paranal Observatory under Program IDs LP181.A-0717,
  LP168.A-0485, ID 170.A-0788, and the ESO Science Archive under
  Program IDs 64.O-0643, 66.A-0572, 68.A-0544, 164.O-0561, 163.N-0210,
  and 60.A-9120.

\end{acknowledgements}

\bibliographystyle{aa}

\end{document}